\documentclass[showpacs,aps,pre,floatfix,amsmath,amssymb]{revtex4}
\usepackage{graphicx}
\usepackage{pstricks}
\usepackage{mathtools}
\begin{document}

\title{Metastable and scaling regimes of a one-dimensional Kawasaki dynamics}

\author{F.\,A. G\'omez Albarrac\'in} 
\author{H.\,D. Rosales}
\author{M.\,D. Grynberg} 
 
\affiliation{Departamento de F\'{\i}sica, Universidad Nacional de  
La Plata, (1900) La Plata, Argentina}

\begin{abstract}
We investigate the large-time scaling regimes arising from a variety of metastable 
structures in a chain of Ising spins with both first- and second-neighbor couplings 
while subject to a Kawasaki dynamics. Depending on the ratio and sign of these 
former, different dynamic exponents are suggested by finite-size scaling analyses
of relaxation times. At low but nonzero-temperatures these are calculated via exact 
diagonalizations of the evolution operator in finite chains under several activation 
barriers. In the absence of metastability the dynamics is always diffusive.
\end{abstract}

	
\maketitle

\section{Introduction}

Systems whose thermodynamic parameters are drastically changed give rise to highly 
nonlinear and far from equilibrium processes that have been intensively studied for 
decades in various contexts \cite{Puri}. These may range from binary fluids to alloys
and spin systems exhibiting a disordered phase at high temperatures, whilst having 
two or more ordered phases below a critical point  \cite{Puri, Puri2}. There is already a 
vast body of research studying such nonequilibrium problems in terms of kinetic Ising 
models \cite{Puri, Puri2, Kawasaki} under both Glauber and Kawasaki dynamics
\cite{Glauber, Kawasaki2}, respectively associated with the so called models $A$ and 
$B$ in the terminology of critical dynamic theories \cite{Hohenberg}. After a quench to
a subcritical temperature these dynamics attempt to minimize the interfacial energy 
between different equilibrium domains which therefore grow and coarsen with time. At 
late stages the spreading of domains is such that if their typical lengths are rescaled by 
a factor $t^{-1/z}$ the domain patterns at different times will be statistically similar. 
Here, the scaling or dynamic exponent $z$ is characteristic of the universality class to 
which the dynamic belongs, and is usually independent of the spatial dimensionality 
but sensitive to conservation laws (see, e.g.  \cite{Puri, Puri2, Hohenberg, Redner} and 
references therein). 

For one-dimensional (1D) systems these dynamics are also amenable to experimental 
probe \cite{Privman}. In particular, the magnetic relaxation of synthesized molecular
chains with strong Ising anisotropy \cite{exp} was considered in the framework of a 
Glauber dynamics with both first- and second-neighbor interactions $J_1, J_2$ \cite{Pini}. 
In that regard, recently it was suggested that weak competing regimes in the low but 
nonzero-temperature limit ($T = 0^+$) give rise to an almost ballistic dynamic exponent 
($z \simeq 1$) \cite{Grynberg} rather than the usual diffusive value of $z = 2$ 
\cite{Puri, Glauber}. Irrespective of how small the frustration might be, note that for 
$0 < -J_2 < \vert J_1\vert$ such discontinuous scaling behavior is also accompanied by 
the sudden appearance of a large basin of metastable states \cite{Redner2}. When it 
comes to the ferromagnetic Kawasaki dynamics, these latter already exist for $J_2 = 0$ 
and are characterized by kinks or domain walls separated by two or more lattice spacings 
\cite{Robin, Redner}. In that situation the mean density of kinks reaches a finite value 
\cite{details}, and so the average size of metastable domains can not but remain 
bounded. At zero temperature the dynamics rapidly gets stuck in these states\, -thus 
preventing the system to reach equilibrium-\, but for $T = 0^+$ their structure is at the 
origin of the $t^{1/3}\!$ growth of domain length scales \cite{Robin, details}. Despite 
metastability, note that as long as temperature is held finite, no matter how small, the 
dynamics is still ergodic and eventually the equilibrium state is accessible. 

Continuing the development initiated in Ref.\,\cite{Grynberg}, in this work we further 
investigate the metastable effects brought about by second neighbor couplings on the 
scaling regimes of this typical phase separation dynamics. As we shall see, for $J_2 \ne 
0$ there are new metastable scenarios and activation barriers that come into play, 
ultimately affecting the large time kinetics in the low temperature limit. 

Following the methodology of Ref.\,\cite{Grynberg}, first  we will construct and 
diagonalize numerically the kink evolution operator associated to the master equation 
\cite{Kampen} of this dynamics in finite chains. This will enable us to determine the 
relaxation time ($\tau$) of these processes from the spectral gaps of that Liouvillian 
operator. As is known, in nearing a critical point those time scales diverge with the 
equilibrium correlation length as $\tau \propto \xi^z$ \cite{Hohenberg}, which in the 
present context also grows unbounded in the limit of $T \to 0^+$ (except for $J_2 = -
\vert J_1 \vert/2\,$ where the ground state is highly degenerate and $\xi$ remains finite 
\cite{Tanaka}\,). On the other hand, in a system of typical length $L$ evidently $\xi$ 
can not grow beyond that scale; hence in practice it is customary to trade that scaling 
relation by the finite-size behavior 
\begin{equation}
\label{scaling}
\tau \propto L^z,
\end{equation}
provided $L$ is taken sufficiently large \cite{Henkel}. Thus, once armed with the above 
referred spectral gaps we will aim at obtaining dynamic exponents from this finite-size 
scaling relation across a host of metastable situations. At low but nonzero-temperatures, 
however notice that owing to the presence of Arrhenius activation energies the time 
scales involved are unbounded even for finite systems (cf. Sec.\,IV). 

The structure of this work is organized as follows. In Sec.\,II we recast the master 
equation of these stochastic processes in terms of a quantum spin analogy that readily 
lends itself to evaluate the low lying levels of the associated evolution operator. Owing to 
detailed balance \cite{Kampen} this latter can be brought to a symmetric representation 
by means of simple nonunitary spin transformations, thus simplifying the subsequent 
numerical analysis. Sec.\,III describes various metastable regimes while attempting to 
identify their decay patterns and activation barriers. Calculational details regarding the 
proliferation rates of these states with the system size are relegated to Appendix A. 
In Sec.\,IV we evaluate numerically the spectrum gaps of finite chains using standard  
recursive techniques \cite{Lanczos} which yield clear Arrhenius trends for relaxation 
times at low temperature regimes. This provides a sequence of finite-size approximants 
to dynamic exponents which are then combined with extrapolations to the
thermodynamic limit \cite{Henkel, Guttmann}. Finally, Sec.\,V contains a summarizing 
discussion along with some remarks on open issues and prospects of future work. 

\vskip 1cm

\section{Dynamics and operators}

Let us consider the Kawasaki dynamics in a chain of Ising spins ($S = \pm 1$) coupled 
with both first- and second-neighbor interactions $J_1, J_2$, thus setting energy 
configurations
\begin{equation}
\label{energy}
E_S = - J_1 \sum_i S_i \, S_{i+1} - J_2 \sum_i S_i \, S_{i+2}\,,
\end{equation}
while in contact with a heat bath at temperature $T$. Here, frustration arises when 
combining antiferromagnetic (AF) $J_2$ couplings ($J_2 < 0$) with $J_1$ exchanges 
of any sign. In particular, for $-J_2/\vert J_1 \vert > 1/2$ the ground state consists of 
consecutive pairs of oppositely oriented spins ($\cdots \bullet \bullet \circ \circ \cdots$), 
while for $0 \le -J_2/\vert J_1 \vert < 1/2$ the ordering is F or AF depending on the 
respective sign of $J_1$. For $J_2 > 0$ there is no frustration and the order type is 
also set by $J_1$. Unless otherwise stated, periodic boundary conditions (PBC) and
a vanishing magnetization will be assumed throughout. 

The bath is represented as causing the Ising states $\vert S\,\rangle = \vert S_1,\,\dots\,, 
S_L\rangle$ to fluctuate by exchanges of nearest neighbor (NN) spin pairs chosen 
randomly from $L$ locations. To enforce the system to relax towards the Boltzmann 
distribution $P_B(S) \propto e^{-E_S /k_B T}$, the transition probability rates per unit 
time $W ( S \to  S')$ between two configurations $\vert S \rangle, \vert S'\rangle$ (here 
differing just in an exchanged pair of NN spins), are chosen to satisfy the detailed 
balance condition \cite{Kampen}
\begin{equation}
\label{DB}
P_B (S) \,W(S \to S') = P_B (S') \, W(S' \to S)\,,
\end{equation}
(also, see its role by the end of Sec.\,II\,A). However, detailed balance itself cannot 
determine entirely the form of such rates, thus in what follows we take up the 
common choice used in the context of kinetic Ising models, namely \cite{Kubo}
\begin{equation}
\label{choice}
W (S \to S') = \frac{\alpha}{2}\Big\{\,1- \tanh \Big[\,\frac{\beta }{2} \,
\big( E_{S'} - E_S \big) \Big]\,\Big\}\,,
\end{equation}  
where $\alpha^{-1}$ just sets the time scale of the microscopic process, and is hereafter
set to 1. Also, from now on temperatures are measured in energy units, or, equivalently, 
the Boltzmann constant in $\beta \equiv 1/(k_BT )$ is taken equal to unity. In the specific 
case of spin exchanges, say at locations $i, i+1$, clearly from Eq.\,(\ref{energy}) the 
above energy differences reduce to  $\frac{1}{2}\,(S_i - S_{i+1}) \left[\,(J_1 - J_2) \left(\, 
S_{i-1} - S_{i+2} ,\right) + J_2 \left(\,S_{i-2}  - S_{i+3}\,\right)\,\right]$. Thus, after 
introducing the parameters
\begin{subequations}
\begin{eqnarray}
\label{pq}
P &=& 2\, (K_1 - K_2)\,,\,\;\;  Q = 2\, (K_1 - 2 K_2)\,,\\
\label{Cpm}
A_{\pm} &=& \pm \frac{1}{8} \left(\,\tanh\,2K_1 - \tanh\,Q\,\right)
+ \frac{1}{4}  \tanh\, 2K_2 \,,\\
\label{Dpm}
B_{\pm} &=& \pm  \frac{1}{8} \left(\,\tanh\,2K_1 + \tanh\,Q\,\right) 
+  \frac{1}{4} \tanh P \,,
\end{eqnarray}
\end{subequations}
($K_{1,2} \equiv J_{1,2} /T$) and using basic properties of hyperbolic functions, it is a 
straightforward matter to verify that the exchange rates $W (S_i ,\, S_{i+1})$
associated to those locations actually deploy multi-spin terms of the form
\begin{eqnarray}
\nonumber
W \left( S_i,\,S_{i+1} \right) &=& \frac{1}{4}\,\big(1 - S_i \, S_{i+1}\big)\,
 -  \frac{1}{4} \, \big( S_i - S_{i+1}  \big)\\
\label{rates}
&\times& \Big[ \big(\, A_{_+} + A_{_-} \; S_{i-1} \; S_{i+2} \big)\,
\big( S_{i-2} - S_{i+3} \big) + \big( B_{_+} + B_{_-} \;S_{i-2}\; S_{i+3}\big)\, 
\big( S_{i-1} - S_{i+2} \big) \Big]\,,
\end{eqnarray}
which already anticipate the type of many-body interactions to be found later on in 
the evolution operator of Sec.\,II\,A. Here, the factors $(1 - S_i \, S_{i+1})$ and $(S_i - 
S_{i+1})$ ensure vanishing rates for parallel spins while contributing to reproduce 
Eq.\,(\ref{choice}) for antiparallel ones. Also, note that although the standard case of 
$J_2 = 0$ is left with terms of just two-spin interactions, its dynamics is {\it not} yet 
amenable for exact analytic treatments at $T > 0$ \cite{Godreche}.

The range of the sites involved in Eq.\,(\ref{rates}), or equivalently, in the energy 
differences of Eq.\,(\ref{choice}), basically distinguishes between eight situations of 
spin exchanges. For later convenience we now regroup them in two sets of dual events 
in which kinks or ferromagnetic domain walls are thought of as hard core particles $A$ 
undergoing pairing $\emptyset + A + \emptyset\,\rightleftarrows\,A + A + A$, and 
diffusion $A + A + \emptyset\,  \rightleftarrows\,\emptyset + A + A$ processes, as 
schematized in Table~\ref{tab1}. Its columns also summarize the information needed 
to construct the operational form of the dynamics, while allowing to infer the variety of 
metastable structures alluded to in Sec.\,I. In what follows we turn to the first of these 
issues, and defer the discussion of the second to Sec.\,III.
\begin{table}[htbp]
\vskip 0.5cm
\begin{center}
\begin{tabular} {c  c  c  c  c}
 \hline \hline
Pairing\;\;\; $\emptyset + A + \emptyset \;\,   \rightleftarrows \;  A + A + A$ &  
\hskip 0.75cm Rate $( \rightleftarrows)$ & \hskip 0.7cm $\beta \Delta E\, 
(\rightleftarrows)$ & \hskip 0.7cm S element & \hskip 0.5cm Projector
\vspace{0.05cm}
\\  
 \hline \hline
\vspace{-0.3cm} \\


$\bullet\; \bullet\; {\blue \bullet\, \vert\, \circ}\, \circ\, \circ \;\;\;
\rightleftarrows \;\;\, \bullet\; \bullet {\blue \vert \circ \vert \bullet \vert} \circ\; \circ$  
& \hskip 0.75cm  $\frac{1}{2} (1\mp \tanh 2 K_{_{\!1}} )$
&  \hskip 0.75cm  $\pm 4 K_{_{\!1}}$ 
&  \hskip 0.75cm  $\frac{1}{2}\,{\rm sech}\,2 K_{_{\!1}}$ 
&  \hskip 0.5cm ${\cal P}^{(1)}$
\vspace{0.3cm}\\

$\circ\, \vert \bullet\, {\blue \bullet\, \vert\, \circ}\, \circ\, \circ \;\;\;
\rightleftarrows \;\;\, \circ\, \vert \bullet {\blue \vert \circ \vert \bullet \vert}\circ\, \circ$
& \hskip 0.75cm   $\frac{1}{2} (1 \mp \tanh P)$
&  \hskip 0.75cm  $\pm 2 P$
&  \hskip 0.75cm  $\frac{1}{2}\,{\rm sech}\,P$
&  \hskip 0.5cm ${\cal P}^{(2)}$
\vspace{0.3cm}\\

$\,\bullet\; \bullet\; {\blue \bullet\, \vert\, \circ\,} \circ \vert\, \bullet \;\;\,
\rightleftarrows \;\;  \bullet\; \bullet {\blue \vert \circ \vert \bullet \vert}
\circ \vert\, \bullet$  
& \hskip 0.75cm   $\frac{1}{2} (1\mp \tanh P)$
&  \hskip 0.75cm  $\pm  2 P$
&  \hskip 0.75cm  $\frac{1}{2}\,{\rm sech}\,P$
&  \hskip 0.5cm ${\cal P}^{(3)}$
\vspace{0.3cm}\\

$\,\circ\, \vert \bullet\, {\blue \bullet\, \vert\, \circ}\, \circ \vert\, \bullet \;\;\,
\rightleftarrows \;\;  \circ\, \vert \bullet {\blue \vert \circ \vert \bullet \vert}
\circ \vert \,\bullet$  
& \hskip 0.75cm   $\frac{1}{2} (1 \mp \tanh Q)$
&  \hskip 0.75cm  $\pm 2 Q$  
&  \hskip 0.75cm  $\frac{1}{2}\,{\rm sech}\,Q$
&  \hskip 0.5cm ${\cal P}^{(4)}$
\vspace{0.3cm}\\


\hline \hline
Diffusion\;\;\, $A + A + \emptyset \;\,   \rightleftarrows \;  \emptyset + A + A$
&   \hskip 0.75cm Rate $( \rightleftarrows)$ & \hskip 0.7cm  
$\beta \Delta E\, ( \rightleftarrows)$ & \hskip 0.7cm S element
& \hskip 0.5cm Projector
\vspace{0.05cm}
\\   \hline \hline
\vspace{-0.3cm} \\

$\circ\; \circ {\blue \vert \bullet \vert\; \circ}\, \circ\, \circ  \;\;\;
\rightleftarrows  \;\; \circ\; \circ\; {\blue \circ\, \vert \bullet \vert}\, \circ \; \circ$  
& \hskip 0.75cm $ 1/2 $ 
&  \hskip 0.75cm  $ 0 $
&  \hskip 0.75cm  $ 1/2 $
&  \hskip 0.5cm ${\cal P}^{(1)}$
\vspace{0.3cm}\\

$\bullet\; \vert \circ {\blue \vert \bullet \vert\, \circ}\, \circ\, \circ  \;\;\,
\rightleftarrows \;\; \bullet\; \vert \circ\, {\blue \circ\, \vert \bullet \vert} \circ\; \circ$  
& \hskip 0.75cm   $\frac{1}{2} (1 \mp \tanh 2 K_{_{\!2}})$
&  \hskip 0.75cm  $\pm 4 K_{_{\!2}}$
&  \hskip 0.75cm  $\frac{1}{2}\,{\rm sech}\,2 K_{_{\!2}}$
&  \hskip 0.5cm ${\cal P}^{(2)}$
\vspace{0.3cm}\\

$\,\circ\; \circ {\blue \vert \bullet \vert\, \circ}\, \circ \vert\, \bullet  \;\;\,
\rightleftarrows  \;\; \circ\; \circ\; {\blue \circ\;  \vert \bullet \vert} \circ \vert\; \bullet$  
& \hskip 0.75cm   $\frac{1}{2} (1\pm \tanh 2 K_{_{\!2}})$
&  \hskip 0.75cm  $\mp 4 K_{_{\!2}}$
&  \hskip 0.75cm  $\frac{1}{2}\,{\rm sech}\,2 K_{_{\!2}}$
&  \hskip 0.5cm ${\cal P}^{(3)}$
\vspace{0.3cm}\\

$\,\bullet\, \vert \circ {\blue \vert \bullet \vert\, \circ}\, \circ \vert\, \bullet \;\;
\rightleftarrows \;\; \bullet\, \vert \circ\, {\blue \circ\; \vert \bullet \vert} \circ 
\vert\; \bullet$  
& \hskip 0.75cm $ 1/2 $ 
&  \hskip 0.75cm  $ 0 $
&  \hskip 0.75cm  $ 1/2 $
&  \hskip 0.5cm ${\cal P}^{(4)}$
\vspace{0.3cm}\\

\hline \hline

\end{tabular}
\end{center}
\vskip 0.3cm
\caption{(Color online) Kawasaki transition probabilities, energy changes, and 
symmetrized (S) non-diagonal matrix elements of the evolution operator transformed 
as in Eq.\,(\ref{s-rates}), for both kink pairing and diffusion processes under $J_1$ 
and $J_2$ interactions. Filled and empty circles denote original spins with opposite 
orientations in turn conforming kinks (vertical lines) on the dual chain. Upper and lower 
signs stand respectively for the forward ($\rightarrow$) and backward ($\leftarrow$) 
processes brought about by exchanging NN spins around central kinks. All events are 
classified according to the projector types defined in Eq.\,(\ref{proj}).}
\label{tab1}
\end{table}

\vspace{-0.28cm}

\subsection{Quantum spin representation}

As is known, in a continuous time description of these Markovian processes the  
stochastic dynamics is controlled by a gain-loss relation customarily termed as the 
master equation \cite{Kampen}
\begin{equation}
\label{MEQ}
\partial_t \,P(S,t) = \sum_{S'}\,\left[\:W (S'\to S)\, P(S',t)\, - \,W (S \to S')\,P(S,t)\:\right]\,,
\end{equation}
which governs the time development of the probability distribution $P(S,t)$.  
Conveniently, this relation can also be reinterpreted as a Schr\"odinger equation in 
imaginary time, i.e.  $\partial_t  \vert P(t)\, \rangle  = - H\, \vert P(t)\, \rangle$ under 
a pseudo-Hamiltonian or evolution operator $H$. This is readily set up by defining  
diagonal and nondiagonal matrix elements \cite{Kawasaki, Kampen}
\begin{subequations}
\begin{eqnarray}
\label{diag}
\langle\,S\,\vert\, H_d\,\vert\,S\,\rangle &=& \sum_{S'\ne S}\, W(S \to S')\,,\\
\label{non-diag}
\langle\,S'\,\vert\,H_{nd}\,\vert\,S\,\rangle &=& -\,W(S \to S')\,,
\end{eqnarray}
\end{subequations}
thus formally enabling to derive the state of the system $\vert P(t)\,\rangle \equiv 
\sum_S  P(S,t) \,\vert S\, \rangle$  at subsequent times from the action of $H$ on a 
given initial condition, that is $\vert P(t) \,\rangle = e^{- H\,t} \vert P(0)\,\rangle$. In 
particular the relaxation time $\tau$ of any observable with nonzero matrix element 
between the steady state and the first excitation mode of $H$, is singled out by the 
eigenvalue  $\lambda_1$ corresponding to that latter, i.e. $1/\tau = {\rm Re}\,
\lambda_1 > 0$, whereas by construction the former merely yields an eigenvalue 
$\lambda_0 = 0$ \cite{Kampen}. Note that the numerical analysis of these spectral
gaps (or inverse relaxation times) will first require to obtain an operational analog 
of Eqs.\,(\ref{diag}) and (\ref{non-diag}), as the phase-space dimension of these 
processes grows exponentially with the system size. That will allow us to implement
the recursive diagonalization techniques of Sec.\,IV, where the matrix representation 
of $H$ is not actually stored in memory \cite{Lanczos}.
 
On the other hand, to halve the number of machine operations it is convenient here to 
turn to a dual description in which new Ising variables $\sigma_i \equiv - S_i\,S_{i+1}$ 
standing on dual chain locations denote the presence ($+1$) or absence ($-1$) of the 
kinks referred to above. Thus, if we think of the states $\vert\, \sigma_1,\,\dots\,\sigma_L 
\rangle$ as representing configurations of $\frac{1}{2}$-spinors (say in the $z$ 
direction), we can readily construct the counterpart of the above matrix elements by 
means of usual raising and lowering operators $\sigma^+,\,\sigma^-\!$. Clearly, the 
nondiagonal parts $H_{nd}^{^{(pair)}}\!\!\!,\,  H_{nd}^{^{(di\!f\!f)}}$ accounting 
for the kink pairing and diffusion processes depicted in Table~\ref{tab1} must involve 
respectively terms of the form $\sigma^{\pm}_{i-1}\,\sigma^{\pm}_{i+1}$ and 
$\sigma^{\pm}_{i-1}\, \sigma^{\mp}_{i+1}$, say for events occurring at locations 
$i-1, i+1$ under the presence of a central kink. However, due to the $J_2$-\,couplings,  
note that these terms should also comprise the kink occupation $\hat n \equiv \sigma^+ 
\sigma^- = \frac{1}{2} (1+\sigma^z)$ and vacancy $\hat{\rm v} \equiv 1 - \hat n$ 
numbers of second neighbor sites surrounding that central kink, as these also matter in 
the rate values of Table~\ref{tab1}. Thus, to weight such correlated processes here we 
classify them according to projectors defined as
\begin{eqnarray}
\label{proj}
{\cal \hat P}_i^{(1)}\!\! &=&  \! \hat {\rm v}_{i-2}\; \hat n_i\; \hat {\rm v}_{i+2}\,,\;\;\;
{\cal \hat P}_i^{(2)}=\,\hat n_{i-2}\; \hat n_i\; \hat {\rm v}_{i+2}\,,\\
\nonumber
{\cal \hat P}_i^{(3)}\!\!  &=& \! \hat {\rm v}_{i-2}\; \hat n_i\; \hat n_{i+2}\,,\;\;\,
{\cal \hat P}_i^{(4)}=\,\hat n_{i-2}\; \hat n_i\; \hat n_{i+2}\,,
\end{eqnarray}
to which in turn we assign the variables $\{x_1, \, x_2, \, x_3, \, x_4\}\equiv \{2K_1, P, P, 
Q\}$, and  $\{y_1, \, y_2, \, y_3,\, y_4\} \equiv \{0, 2 K_2, -2 K_2, 0\}$. Therefore, with the 
aid of these latter, the contributions of the pairing and diffusion parts to the operational 
analog of Eq.\,(\ref{non-diag}) can now be written down as
\begin{subequations}
\begin{eqnarray}
\label{pairing}
H_{nd}^{^{(pair)}}&=& -  \sum_{i,\,j}\, {\cal \hat P}_i^{(j)} \big[\,f (x_j)\,\sigma
^ -_{i-1}\,\sigma^-_{i+1} + f (-x_j)\,\sigma^+_{i+1}\,\sigma^+_{i-1}\,\big]\,,\\
\label{diffusion}
H_{nd}^{^{(di\!f\!f)}} &=& -  \sum_{i,\,j}\,{\cal \hat P}_i^{(j)} \big[\,f (y_j)\, 
\sigma^+_{i-1}\,\sigma^-_{i+1} + f (-y_j)\,\sigma^+_{i+1}\,\sigma^-_{i-1}\,\big]\,,
\end{eqnarray}
\end{subequations}
where $f (u) \equiv \frac{1}{2} (1 + \tanh u)$, and the $j$ index runs over the four 
types of projectors specified in Eq.\,(\ref{proj}).

When it comes to the diagonal terms associated with Eq.\,(\ref{diag}), in turn needed 
for conservation of probability, notice that these basically count the number of manners 
in which a given configuration can evolve to different ones in a single step. In the kink 
representation this amounts to the summation of all pairing and diffusion attempts that 
a given state is capable of. As before, these attempts also can be probed and weighted 
by means of the above projectors, vacancy, and number operators, in terms of which 
those diagonal contributions are expressed here as
\begin{subequations}
\begin{eqnarray}
H_d^{^{(pair)}}\!\!\!\! &=&  \sum_{i,\,j}\,{\cal \hat P}_i^{(j)} \big[\,f (-x_j)\, \hat 
n_{i-1}\,\hat n_{i+1} + f (x_j)\,\hat{\rm v}_{i-1}\, \hat{\rm v}_{i+1} \,\big]\,,\\
H_d^{^{(diff)}}\!\!\!\! &=&   \sum_{i,\,j}\,{\cal \hat P}_i^{(j)} \big[\,f (-y_j)\, 
\hat{\rm v}_{i-1}\, \hat n_{i+1} +  f (y_j)\,\hat n_{i-1}\, \hat{\rm v}_{i+1}\,\big]\,.
\end{eqnarray}
\end{subequations}
Thus, after simple algebraic steps and using the $A_{\pm},\,B_{\pm}$ 
parameters defined in Eqs.\,(\ref{Cpm}) and (\ref{Dpm}), the net contribution 
${\cal H}_d = H_d^{^ {(pair)}}\!\!\!\!+ H_d^{^{(di\!f\!f)}}\!\!$ of these diagonal 
terms is found to involve two-, three-, and four-\,body interactions of the form
\begin{eqnarray}
\nonumber
{\cal H}_d = \frac{1}{4}\,\sum_i\,  \left(1+\sigma^z_i\right) \!\!\!\! &\Big[& \!\!\!1 + 
\left(\,B_{_+} - B_{_-}\;\sigma^z_{i-2}\, \sigma^z_{i+2}\, \right)
\left(\sigma^z_{i-1} + \sigma^z_{i+1} \right) \\
\label{diagonal}
&-&\! \left(  A_{_+}\, \sigma^z_{i-1} -  A_{_-}\, \sigma^z_{i+1}\right) 
\sigma^z_{i-2}\,+\left( A_{_-}\, \sigma^z_{i-1} - A_{_+}\, 
\sigma^z_{i+1}\right) \sigma^z_{i+2}\, \Big]\,,
\end{eqnarray}
some of which had already appeared at the level of the original spin rates mentioned
in Eq.\,(\ref{rates}).

{\it Detailed balance}.-- Further to the correlated pairing and diffusion terms of 
Eqs.\,(\ref{pairing}),\,(\ref{diffusion}) which would leave us with a non-symmetric 
representation of the evolution operator, we can make  some progress  here by 
exploiting detailed balance [\,Eq.\,(\ref{DB})\,]. This latter warrants the existence 
of representations in which $H$ is symmetric and thereby fully diagonalizable
\cite{Kampen}. For our purposes, it suffices to consider the {\it diagonal} 
nonunitary similarity transformation
\begin{equation}
\label{transf}
\mathbb T = \exp \left[\,\frac{1}{2} \sum_i \big(\, K_1\, \sigma^z_i \,+\,
K_2 \,\sigma^z_i \sigma^z_{i+1}  \big)\,\right]\,,
\end{equation}
stemming from the original spin energies of Eq.\,(\ref{energy}) but re-expressed 
in terms of kinks, i.e. $E_{\sigma} = \sum_i \left( J_1\, \sigma_i + J_2\, \sigma_i\,
\sigma_{i+1} \right)$. Hence $\mathbb T\,\vert \sigma \rangle = e^{\frac{\beta}{2} 
E_{\sigma}} \vert \sigma \rangle$, implying that the nondiagonal matrix elements of 
$\mathbb T\, H\, \mathbb T^{-1}$ will transform as
\begin{equation}
\label{s-rates}
W ( \sigma \to \sigma') \to e^{\,{\frac{\beta}{2}}\, \left(E_{\sigma'} - E_{\sigma}
\right)}\, W ( \sigma \to \sigma')\,.
\end{equation}
But since in the kink representation $W ( \sigma \to \sigma')$ also comply with the 
detailed balance condition (\ref{DB}), then clearly these elements become symmetric 
under $\mathbb T$ (see the symmetrized elements of Table~\ref{tab1}). 
Equivalently, under Eq.\,(\ref{transf}) the pairing and diffusion operators involved in
Eqs.\,(\ref{pairing}) and (\ref{diffusion}) will transform respectively as 
\begin{subequations}
\begin{eqnarray}
\phantom{\Big[}
 \sigma^{\pm}_{i-1} \,\sigma^{\pm}_{i+1}  & \to & 
\exp \Big[  \pm K_2 \left ( \sigma^z_{i-2} + 2 \sigma^z_i + \sigma^z_{i+2}\right) 
\pm 2 K_1 \Big]\; \sigma^{\pm}_{i-1} \,\sigma^{\pm}_{i+1}\;,\\
\phantom{\Big[}
\sigma^{\pm}_{i-1} \,\sigma^{\mp}_{i+1}  & \to &
\exp \Big[  \pm K_2 \left (   \sigma^z_{i-2} -  \sigma^z_{i+2}\right) \Big]\;
\sigma^{\pm}_{i-1} \,\sigma^{\mp}_{i+1}\;,
\end{eqnarray}
\end{subequations}
while leaving ${\cal H}_d$ and all projectors of Eq.\,(\ref{proj}) unchanged. Thus, after 
introducing the $C_{\pm}$ and $D$ coefficients
\begin{subequations}
\begin{eqnarray}
C_{\pm} &=& \frac{1}{2} \left(\,{\rm sech}\,Q + {\rm sech}\,2K_1\,\right)
 \pm {\rm sech}\, P \;,\\
D &=&  \frac{1}{2} \left(\,{\rm sech}\,Q - {\rm sech}\,2K_1\,\right) \,,
\end{eqnarray}
\end{subequations}
it is straightforward to check that the symmetric counterparts of 
$H_{nd}^{^{(pair)}}\!\!$ and  $H_{nd}^{^{(di\!f\!f)}}\!$ are then given by
\begin{subequations}
\begin{eqnarray}
\label{s-pairing}
{\cal H}_{nd}^{^{(pair)}} & = & - \frac{1}{8}\,\sum_i\,  \left(1+\sigma^z_i\right)
\Big[\, C_{_{\!+}} + D \left(\,\sigma^z_{i-2} + \sigma^z_{i+2}\right) +  
C_{_{\!-}}\,\sigma^z_{i-2} \, \sigma^z_{i+2} \,\Big] \,
\left(\,\sigma^+_{i-1}\,\sigma^+_{i+1} + {\rm H.c.}\,\right)\,,\\
\label{s-diffusion}
{\cal H}_{nd}^{^{(di\!f\!f)}} & = & - \frac{1}{8}\,\left(\,1 + {\rm sech}\, 2K_2 \,
\right)\,\sum_i \,\left(1+\sigma^z_i\right)  \left(\,1+\tanh^2\!K_2\; \sigma^z_{i-2}\, 
\sigma^z_{i+2}\,\right) \,\left(\,\sigma^+_{i-1}\,\sigma^-_{i+1} + {\rm H.c.}\,\right)\,.
\end{eqnarray}
\end{subequations}

Together with Eq.\,(\ref{diagonal}) this completes the construction of the operational
analog of Eqs.\,(\ref{diag}) and (\ref{non-diag}) in an Hermitian representation. In 
passing, it is worth to point out that all above non-diagonal operators not only preserve 
the parity of kinks $e^{i \pi \sum_j \hat n_j}$ (being even for PBC), but that they also 
commute, by construction, with $\sum_j  e^{i \pi \sum_{k < j}\hat n_k},$ which simply  
re-expresses the conservation of the total spin magnetization in the original system. In 
practice, for the numerical evaluation of spectral gaps (Sec.\,IV) we will just  build up the 
adequate basis of kink states from the corresponding spin ones.

\section{Metastable states}

After an instantaneous quench down to a low but non-zero temperature often this 
stochastic dynamics rapidly reaches a state in which further energy-lowering processes 
are unlikely. This is because the configuration space contains `basins' of local energy 
minima from which the chances to access lower energy states must first find their way 
through a typical `energy barrier' $E_b$. In the limit of $T \to 0^+$ the average time 
spent in these configuration, or metastable (M) states, then diverges with an Arrhenius 
factor $e^{\beta E_b}$. In common with the standard ferromagnetic dynamics, here the 
decay from these M-\,configurations is mediated by diffusion of kink pairs. In particular, 
for $J_2 = 0$ their release requires activation energies of $4 J_1$ which at 
low-temperatures involves time scales $\propto e^{4 K_1}$ (see pairing rates of 
Table~\ref{tab1}). As a result of the diffusion of these pairs, entire ferromagnetic 
domains can move rigidly by one lattice spacing \cite{details}. Following an  
argumentation given in Refs.\,\cite{Redner, Robin}, the repeated effect of that rather 
long process ultimately leads to coarsening of domains, and is at the root of their 
$t^{1/3}$ growth \cite{details}.

For $J_2 \ne 0$ however there are other energy barriers that also come into play, 
so the identification of a net Arrhenius factor in the actual relaxation time is less 
straightforward. In addition, due to the discontinuities already appearing at the level 
of transition rates (specifically at $J_2/J_1= 0, 1/2, 1$ in the limit of $T \to 0^+$), note 
that there are several coupling regimes where that identification must be carried out. As 
we shall see, that will prove much helpful in the finite-size scaling analysis of  Sec.\,IV 
thus, here we focus attention on the variety of M-\,structures arising in the coupling 
sectors {\it (a)} to {\it (h)} listed in Table~\ref{tab2}. Reasoning with Table~\ref{tab1} 
and guided by simulated quenches down to $T = 0$, we turn to the characterization of 
these structures while trying to identify their decay patterns. Also, a measure of the 
basin of these states, such as the rate at which they proliferate with the system size, is 
provided with the aid of Appendix A. The results of the arguments and observations that 
follow in this Section are summarized in Table~\ref{tab2}.

\begin{table}[htbp]
\normalsize
\vskip 0.2cm
\begin{center}
\begin{tabular} {c  c  c  c }
\hline \hline
\hspace{0.5cm}Coupling regime  \hspace{0.5cm} &  \hspace{0.8cm} Typical  M-state
\hspace{0.3cm} & \hspace{0.1cm} Rates ($\sim x^L$)  \hspace{0.3cm} & Barrier 
($\beta\, E_b \propto \ln \tau$)
\vspace{0.05cm}
\\  
\hline \hline
\vspace{-0.2cm} \\
\;\;\;{\it (a)} $\;\;\; J_1, J_2 \ge 0\;,  \;\;\;0 \le r < \frac{1}{2}$ &  \hskip 0.7cm ${\blue\bf 
1}\,\underbracket[0.5pt]{\,0\, \cdots\,}_{\rm v\,\ge 1}\, {\blue\bf 1}\; 0\, \cdots$ & 
\hskip -0.8cm  1.6180 &  $4 \left( K_1 \!+ \!K_2 \right)$ 

\vspace{0.3cm}\\
\;{\it \,(b)} $\;\;\;J_1, J_2 > 0\;,\;\;\;\frac{1}{2} \le r < 1$ &  \hskip 0.7cm ${\blue\bf 1}\, 
\overbracket[0.5pt]{\underbracket[0.5pt]{\,0\, \cdots\,}_{\rm v\, \ge 1}\, {\blue\bf 1}\, 
\underbracket[0.5pt]{\,0\, \cdots\,}_{\rm v'\,\ge 1}}^{\rm v\,\,\,or\,\,\, v'\,>\, 1}\, 
{\blue\bf 1}\, \cdots$ & \hskip -0.8cm  1.5701  & $4 \left( K_1 \!+ \!K_2 \right)$ 

\vspace{0.42cm}\\
\hskip -1.5cm {\it (c)} $\;\;\;\;J_1 = J_2 > 0$ & \hskip 0.6cm ${\blue\bf 1}\, 
\underbracket[0.5pt]{\,0\, \cdots\,}_{\rm v\, \ge 2}\,{\blue\bf 1}\;0\, \cdots$ & 
\hskip -0.8cm 1.4655 & $4 \left( K_1 \!+ \!K_2 \right)$ 

\vspace{0.4cm}\\
\hskip -1.5cm {\it (d)} $\;\;\;\;J_2 > J_1 > 0$ &  \hskip 0.54cm  $\underbracket[0.5pt]
{\,{\blue\bf 1}\,\cdots\,}_{k\, \ne 2,3}\;\underbracket[0.5pt]{\,0\, \cdots\,}_{\rm v\, \ge
2}\, {\blue\bf 1}\,\cdots$ &  \hskip -0.8cm 1.6180 & $4 \left( K_1 \!+ \!K_2 \right)$ 

\vspace{0.55cm}\\
\hskip -1cm {\it (e)} $\;\;\; J_1 < 0\;,\;\;J_2 > 0$ & \hskip 0.55cm $\underbracket[0.5pt]
{\,{\blue\bf 1}\,\cdots\,}_{k\, \ge 3}\;\underbracket[0.5pt]{\,0\, \cdots\,}_{\rm v\, \ge 2} 
\,{\blue\bf 1}\,\cdots$ & \hskip -0.8cm 1.5289 & $4 K_2$

\vspace{0.55cm}\\
{\it \;(f)} $\;\;\; J_1, J_2 < 0\;, \;\;\;\frac{1}{2} < r \le 1$ &  \hskip 0.54cm 
$\underbracket[0.5pt]{\,{\blue\bf 1}\,\cdots\,}_{k\,=1,2}\,0\; {\blue\bf 1}\,\cdots$
& \hskip -0.8cm 1.3247 & \hskip -0.2cm  $\!\!-4 K_2$

\vspace{0.35cm}\\
\hskip -1.6cm {\it (g)} $\;\;\;J_2 < J_1 < 0$ &    \hskip 1.1cm $\overbracket[0.5pt]
{\underbracket[0.5pt]{\,0\, \cdots\,}_{\rm v\, \ge 1}\; \underbracket[0.5pt]
{\,{\blue\bf 1}\,\cdots\,}_{k\,=1,2} \; \underbracket[0.5pt]{\,0\, \cdots\,}_{\rm v'\,
\ge 1}}^{ {\rm v \,\,\, or \,\,v' =\, 1,\,\,\,if} \,\,\, k\,=1}\,{\blue\bf 1}\, \cdots$
& \hskip -0.8cm  1.7437  &  $\;4 \left( K_1 \! -  2  K_2 \right)$

\vspace{0.45cm}\\
\hskip -1.08cm {\it (h)} $\;\;\;    J_1 > 0\;,\,J_2 < 0$
&  \hskip 0.57cm $\underbracket[0.5pt]{\,{\blue\bf 1}\,\cdots\,}_{k\,=1,2}\,
\underbracket[0.5pt]{\,0\, \cdots\,}_{\rm v\, \ge 1}\,{\blue\bf 1}\,\cdots$ & 
\hskip -0.8cm 1.8392 & $\;4 \left( K_1 \! -  2 K_2 \right)$

\vspace{0.3cm}\\
\hline \hline 

\end{tabular}
\end{center}
\vskip 0.35cm
\caption{(Color online) Schematic configurations of metastable states (in dual 
representation) for several coupling regimes (here $r \equiv J_2 / J_1$). The former 
are composed by constrained sequences of kinks and vacancies $(k, \rm v)$ as 
indicated by brackets. Proliferation rates of these configurations with the system 
size (calculated in Appendix A), along with activation energy barriers associated to
their relaxation times, are quoted on the rightmost columns. Free diffusion of kink 
pairs ($\Delta E=0$) between jammed sequences may occur in cases {\it (g)} and 
{\it (h)} (see text for details).}
\label{tab2}
\end{table}

\vskip 0.25cm
{\it (a)} and {\it (b)}.--\, In  these two first coupling sectors the kinks of all 
M-\,states must be separated by at least one vacancy because in the pairing rates 
of Table~\ref{tab1}\, both $P$ and $K_1$ are positive. However, note that in sector 
{\it (b)} sequences of the form $...\,1 0 1 0 1\,...$ could not show up because there 
$Q \le 0$ (see fourth pairing process of Table~\ref{tab1}). As is indicated in Appendix 
A this further constraint (schematized in Table~\ref{tab2}), significantly reduces the 
proliferation of M-\,states with respect to sector {\it (a)}. For this latter the number of 
M-\,configurations turns out to grow as $g^L$ with golden mean $g \sim 1.618$, 
whereas for sector for {\it (b)} it grows only as $\sim 1.5701^L$.

As mentioned above, on par with the usual dynamics of $J_2 = 0$ the low-temperature
decay from either of these structures involves the diffusion of kink pairs between 
otherwise isolated domain walls \cite{details}. However, notice that for $J_2 >0$ 
this requires the occurrence of two successive and raising energy events (cf. 
Table~\ref{tab1}) namely, pairing around an existing wall ($\Delta E = 4 J_1$) followed 
by detachment of neighboring pairs from the triplets so formed ($\Delta E = 4 J_2$). 
This allows the kink pair to diffuse at no further energy cost until eventually a new wall 
is encountered and the energy excess is rapidly released \cite{details}. The net thermal 
barrier of this composite process is therefore $ \beta E_b = 4 (K_1 + K_2)$, thus setting 
arbitrarily large relaxation time scales in the low-temperature limit, even for {\it finite} 
lattices. We will find back these time scales later on in the exact diagonalizations of 
Sec.\,IV\,A. Yet, it remains to be determined whether the above reduction of metastability 
(as well as that of the following case), is of any consequence for dynamic exponents.

\vskip 0.25cm
{\it (c)}.--\, Here $Q < 0$ but now $P = 0$, implying that kinks must at least be 
separated by two vacancies. Otherwise, the isolated vacancies involved both in the 
second and third processes of Table~\ref{tab1} would alternately originate a random 
walk of pairings at no energy cost until encountering another kink. Then the walk could 
no longer advance, and eventually after few but energy-\,decreasing processes the 
isolated vacancy would be finally canceled out. This constraint brings about even further 
reductions in the number of M-\,states which actually now grows as $\sim 1.4655^L$
(Appendix A). However, this new minimum separation of kinks has no effect in the decay 
pattern mentioned in the previous two cases, so here the Arrhenius factors can also be 
expected to diverge as $e^{4\,(K_1+K_2)}$. As before, this will be corroborated in 
Sec.\,IV\,A.

\vskip 0.25cm
{\it (d)}.--\, In this coupling regime $P$ and $Q$ are both negative, meaning that groups 
of four or more consecutive kinks, i.e. AF spin domains, may well show up in these new 
M-\,states (see Table~\ref{tab1}). Also, since $K_1 > 0$ isolated kinks can still appear 
scattered throughout. In turn, these latter as well as all AF domains must be separated 
by at least two vacancies otherwise, as indicated in Table~\ref{tab1}, the energy would 
decrease further. On the other hand, it turns out that similarly to sector {\it (a)} the 
number of M-\,states here also proliferates with the golden mean as $g^L$ (see 
Appendix A).

When it comes to energy barriers, at a first  stage on time scales $\tau_{_{\!P}} 
\propto e^{-2 P}$, and so long as $J_2 < 2 J_1$, most AF domains can disgregate by 
pair annihilations followed by energy-\,decreasing processes. The steps involved may 
be schematically represented as (say, starting from the leftmost triplet)
$$\cdots\,0\,0\,1\,1\,1\,1\,1\,1\,\cdots\;^{\Delta E_i \,>\, 0}_{\;\;\longrightarrow}\,
\cdots\,0\,0\,0\,1\,0\,1\,1\,1\,\cdots\; ^{\Delta E\,=\, 0}_{\;\;\longrightarrow}\,\cdots\,
0\,0\,0\,1\,1\,1\,0\,1\,\cdots\; ^{\Delta E_f \,<\, 0}_{\;\;\;\longrightarrow}\,\cdots\, 
0\,0\,0\,0\,1\,0\,0\,1\,\cdots$$ 
where $\beta\, \Delta E_i = - 2P$, and $\beta\, \Delta E_f = - 4 K_1$. (Note that 
annihilations around innermost kinks would require even larger time scales of order 
$e^{-2Q}$ because $\vert Q \vert  > \vert  P \vert\,$  throughout this regime). Since at 
low temperatures $\tau_{_{\!P}} \ll e^{4 K_1}$ the process can recur and eventually 
the entire AF domain can be disaggregated on times $\propto \tau_{_{\!P}}$\,. 
However, for $J_2 \ge 2 J_1$ that process could no longer advance within those time 
scales. Instead, disgregation simply proceeds via successive detachment of outer pairs, 
each one requiring scales $\tau_2 \propto e^{4 K_2} \gg \tau_{_{\!P}}$. In either case, 
these pairs can then realize a free random walk ($\Delta E = 0$), until finally annihilating 
with an isolated kink ($\Delta E = - 4 J_1$). In a much later second stage, when most 
consecutive kinks disappear, large vacancy regions then proceed to coarse-grain as in 
previous cases \cite{details}. As before, this ultimately requires time scales of order 
$e^{4\,(K_1+K_2)} \gg \tau_2,\, \tau_{_{\!P}}$ (cf. with spectral gaps of Sec.\,IV\,A).

\vskip 0.25cm
{\it (e)}.--\, Much alike the previous situation, here groups of consecutive kinks 
separated by two or more vacancies  may also appear in these M-\,sequences 
because $P, \,Q,$ and $-K_2$ are still negative. However, since $K_1 < 0$ there can be
no isolated kinks, and consequently AF groups may now range from three kinks onwards 
(see Table~\ref{tab1}). As is shown in Appendix A, the proliferation of M-\,states given 
rise by these new constraints turns out to grow as $\sim 1.5289^L$.

Like in the initial stages of case {\it (d)}, the decay to equilibrium (now fully AF) 
also proceeds through a random walk of kink pairs which successively detach from 
AF domains on time scales $\propto e^{4 K_2}$ \cite{comment1}. As the random 
walk goes on, these pairs may coalesce with others, or stick back to the same or to a new 
domain (in all cases with $\Delta E \le 0$), until eventually some groups get fragmented 
into triplets \cite{comment2}. Yet, within times $\propto e^{4 K_2}$ these may further 
decompose as 
$$\cdots\,0\,1\,1\,1\,0\,\cdots\;\,^{\Delta E \,=\, 4 J_2}_{\;\;\;\;\,\longrightarrow}\;
\cdots\,0\,1\,0\,1\,1\,\cdots\;\,^{\Delta E \,<\, -4 J_2}_{\;\;\;\;\;\longrightarrow}\;
\cdots\,1\,1\,1\,1\,1\,\cdots$$ 
say, detaching a rightmost pair from one of those triplets. Thus, in the low-temperature 
limit the above process is unlikely to reverse on these time scales, and so the process can 
recur in a progressively denser medium until the full AF state is reached. As we shall see 
in Sec.\,IV\,B this decay pattern, now activated by $4 J_2$ energies, actually brings about 
a faster coarsening than those obtained in previous cases.

\vskip 0.25cm
{\it (f)}.-- As suggested by simulated quenches down to $T = 0$, for vanishing 
magnetization there are no consecutive vacancies in the M-\,states of this 
regime \cite{separations}. Since here $Q > 0$ and $K_2, P < 0$, then following 
Table~\ref{tab1} these configurations should be consistent with single and, at most, 
four contiguous kinks scattered throughout. However, for $S^z = 0$ it turns out that just 
single and double kinks actually show up \cite{separations}. The recursions of Appendix 
A then show that these tighter constraints give rise to a number of M-\,states which 
proliferates only as $\sim 1.3247^L$.

As for the manner in which such configurations proceed to equilibrium, contrary to the 
previous situations it is difficult to identify here a specific pattern of decay. Rather, we 
content ourselves with mentioning that the largest energy barrier of these states (i.e. 
$-4 J_2$ actually corresponding to the activation of diffusing pairs), turns out to be 
associated with the actual relaxation times $\propto e^{- 4 K_2}$ evaluated by the 
exact diagonalizations of Sec.\,IV\,C.

\vskip 0.25cm
{\it (g)} and {\it (h)}.-- For these two last sectors the signs of $Q$ and $K_2$ remain as 
in the previous case, though now $P$ is positive. Thus, following Table~\ref{tab1}, as 
before only single and double kinks may appear distributed throughout but now they can 
be separated by one or more vacancies, even for $S^z= 0$. Thereby, kink pairs are able 
to diffuse freely ($\Delta E = 0$) across consecutive vacancies while keeping at least one 
space from each other and other kinks. So, the typical M-\,state of these sectors actually 
results in an alternating sequence of mobile and jammed blocks. In these latter there 
can be just one vacancy aside each kink pair, whereas single kinks may appear separated 
by one or more spaces. However, note that in sector {\it (g)} sequences of the form 
$...\,0 0 1 0 0\,...$ would be unstable because there $K_1 < 0$ (see first pairing process
of Table~\ref{tab1}). On par with what occurs in cases {\it (a)} and {\it (b)}, this further 
constraint (schematized in Table~\ref{tab2}) then reduces the proliferation of 
M-\,configurations with respect to sector {\it (h)}. In fact, the recursions of Appendix A 
show that for that latter sector the number of M-\,states grows as $\sim 1.8392^L$, 
while for {\it (g)} it grows just as $\sim 1.7437^L$. 

With regard to the decay of these states, as in case {\it (f)} here we limit ourselves 
to mention that in nearing low-temperature regimes it turns out that it is the largest 
activation barrier the one that takes over (fourth paring process of Table~\ref{tab1}), 
as the Arrhenius factors of these sectors actually will turn out to diverge as $e^{2 Q}$ 
(see Sec.\,IV\,C).

\vskip 0.25cm
Finally, notice that in the region $J_1 < 0$ with $0 \le J_2 / J_1 < 1/2$ there are {\it no} 
M-\,basins hindering the access to the AF ground state, i.e. $E_b \equiv 0$. Here, the 
paths to this latter proceed much as in sector {\it (e)} except that now $J_2 < 0$, 
so the dynamics can further decrease the energy either by triplet splittings or pair 
creations ($K_1, P, Q < 0$). Therefore as temperature is lowered in this region, the 
relaxation time towards the AF ordering of {\it finite} chains remains bounded (see 
beginning of Sec.\,IV\,C).

\section{Scaling regimes}

Having built up the kink evolution operator in a symmetric representation 
[\,Eqs.\,(\ref{diagonal}), (\ref{s-pairing}), and (\ref{s-diffusion})\,], next we proceed
to evaluate numerically its spectral gap in finite chains via a recursion type Lanczos 
algorithm \cite{Lanczos}. As mentioned in Sec.\,II we focus on the case of vanishing 
magnetization in the original spin model, thus corresponding to a subspace of 
$\frac{1}{2} \binom {L}{L/2}$ kink states. As a preliminary test first we verified 
that the transformed Boltzmann distribution $\left\vert \psi_0 \right \rangle \propto 
\sum_{\sigma} \exp(- \frac{\beta}{2} E_{\sigma})\, \vert \sigma \rangle$ resulting 
from Eq.\,(\ref{transf}), actually yields the 'ground' state of our quantum `Hamiltonian' 
with eigenvalue $\lambda_0 \equiv 0$. Thereafter, as usual the Lanczos recursion was 
started  with a random linear combination of kink states, but here chosen orthogonal to 
that Boltzmann-like direction. In turn, all subsequent vectors generated by the Lanczos 
algorithm were also reorthogonalized to $\left\vert \psi_0  \right \rangle$. This allowed 
us to obtain the first excited eigenmodes of the evolution operator in periodic chains of 
up to $L = 24$ sites, the main limitation for this being the exponential growth of the 
space dimensionality. 

Another restrictive issue we are confronted with is that as temperature decreases the 
spectral gaps ($\lambda_1$) get arbitrarily small due to the energy barriers mentioned 
in Sec.\,III, i.e. $\lambda_1 = \tau^{-1} \propto e^{-\beta E_b}$. On the other hand, to 
ensure that these finite-size quantities are actually scaled within the Arrhenius regime, 
in this context it is more appropriate to put forward a `normalized' version of the scaling 
hypothesis (\ref{scaling}), namely
\begin{equation}
\label{normalization}
\Lambda_1^{\!^*} (L) \coloneqq \lim_{T \to 0^+}e^{\,\beta E_b} \,\lambda_1 (L,T) =
A\, L^{-z},
\end{equation}
where the amplitude $A$ would involve at most a $J_1, J_2$ dependent quantity 
(assuming $L$ is large enough). Moreover, alike dynamic exponents such proportionality 
factors will also come out as sector-wise universal constants. In practice, below $T/\vert 
J_1 \vert \sim 0.2$ the evaluation of (\ref{normalization}) requires the use of at least 
quadruple precision but as the spacing between low-lying levels gets progressively 
narrow, for $T/ \vert J_1 \vert \alt 0.1$ it turns out that the pace of the Lanczos 
convergence becomes impractical in most of the coupling sectors of Table~\ref{tab2}.
Nevertheless, as we shall see in the following subsections, already when temperature 
is lowered within the ranges in hand the normalized gaps $\Lambda_1 (L,T) \coloneqq 
e^{\,\beta E_b} \,\lambda_1 (L,T)$ exhibit clear saturation trends, thus constituting 
accurate estimations of $\Lambda_1^{\!^*} (L)$.

\subsection{$J_1, \, J_2 > 0$}

\begin{figure}[htbp]
\hskip -0.75cm
\includegraphics[width=0.56\textwidth]{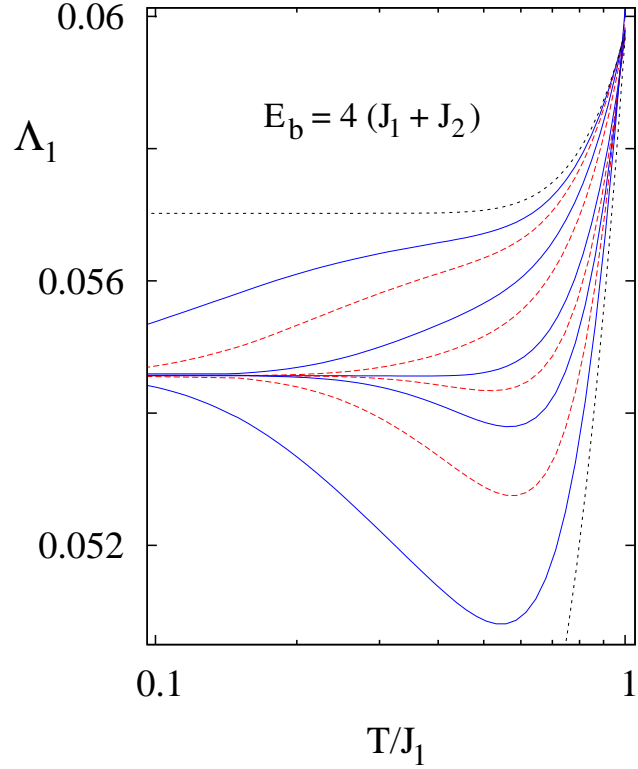}
\vskip 0.25cm
\caption{(Color online) Normalized spectral gaps $\Lambda_1 \equiv e^{E_b/ T}\, 
\lambda_1$ of the evolution operator for $L = 20$ on approaching low-temperature 
regimes in sectors {\it (a)} and {\it (b)} of Table~\ref{tab2}. From top to bottom, 
alternating solid and dashed lines stand for coupling ratios $r \equiv J_2/J_1 = 0.95,\, 
0.9,\, 0.8,\, 0.7,\, 0.5,\, 0.4,\, 0.3,\, 0.2,\, 0.1$. Upper- and lowermost dotted curves 
denoting respectively the cases $r = 1$ [\,sector {\it (c)}\,] and $r=0$, are shown for 
comparison. Details of that latter standard case are displayed in Fig.\,\ref{two}(a).}
\label{one}
\end{figure}

\begin{figure}[htbp]
\vskip 0.1cm
\hskip -13cm
\includegraphics[width=0.37\textwidth]{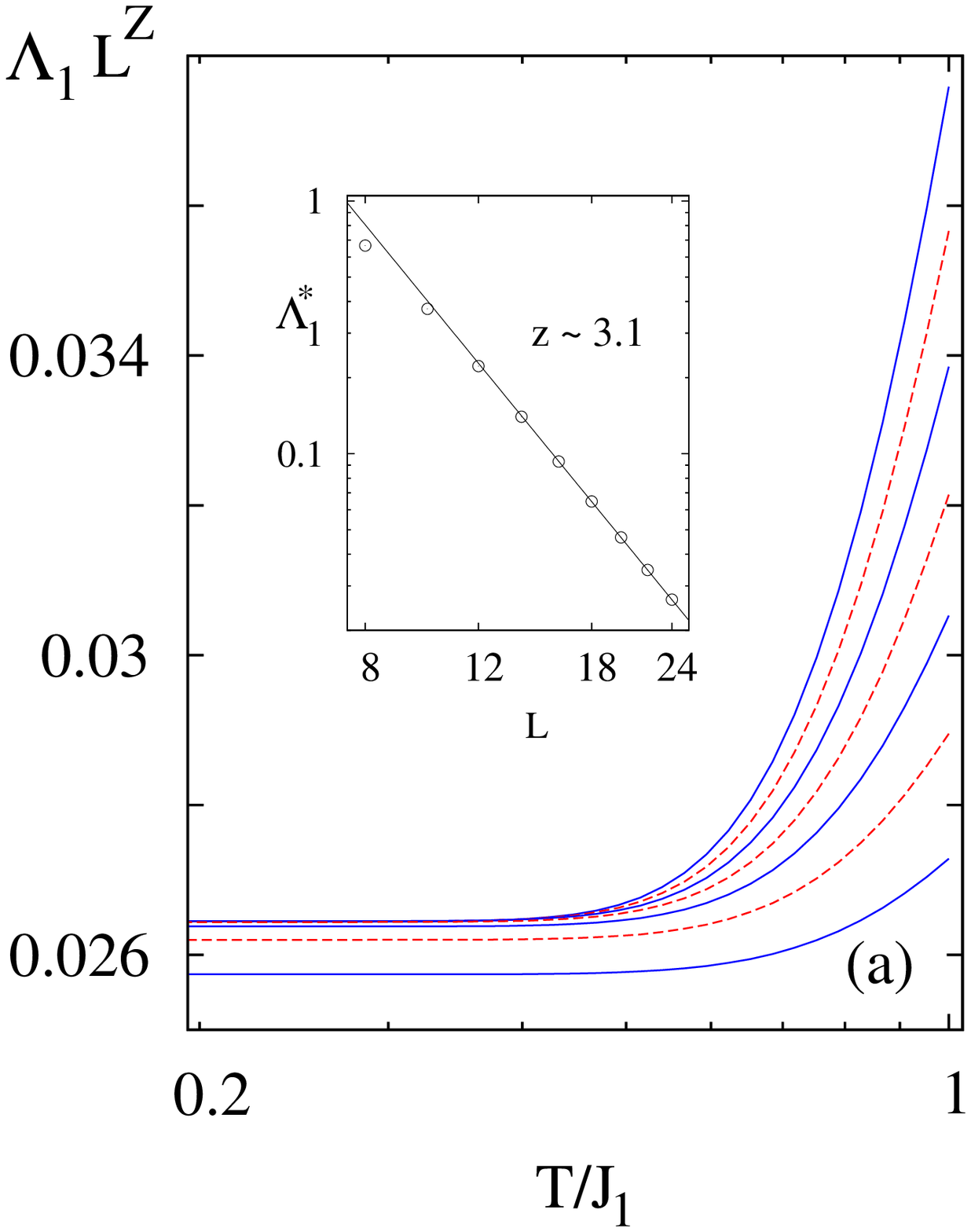}
\vskip -6.8cm
\hskip -1cm
\includegraphics[width=0.37\textwidth]{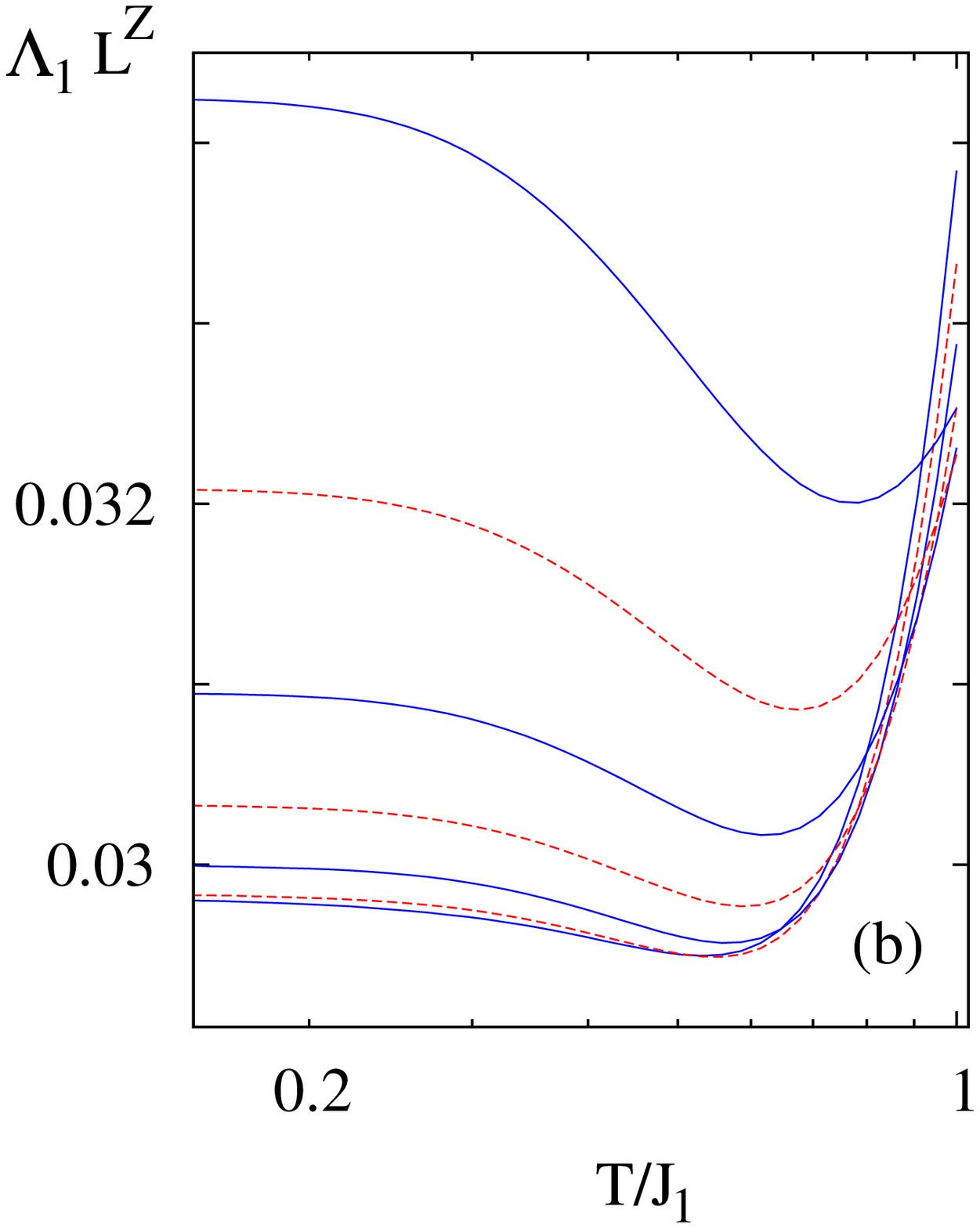}
\vskip -6.82cm
\hskip 11.2cm
\includegraphics[width=0.37\textwidth]{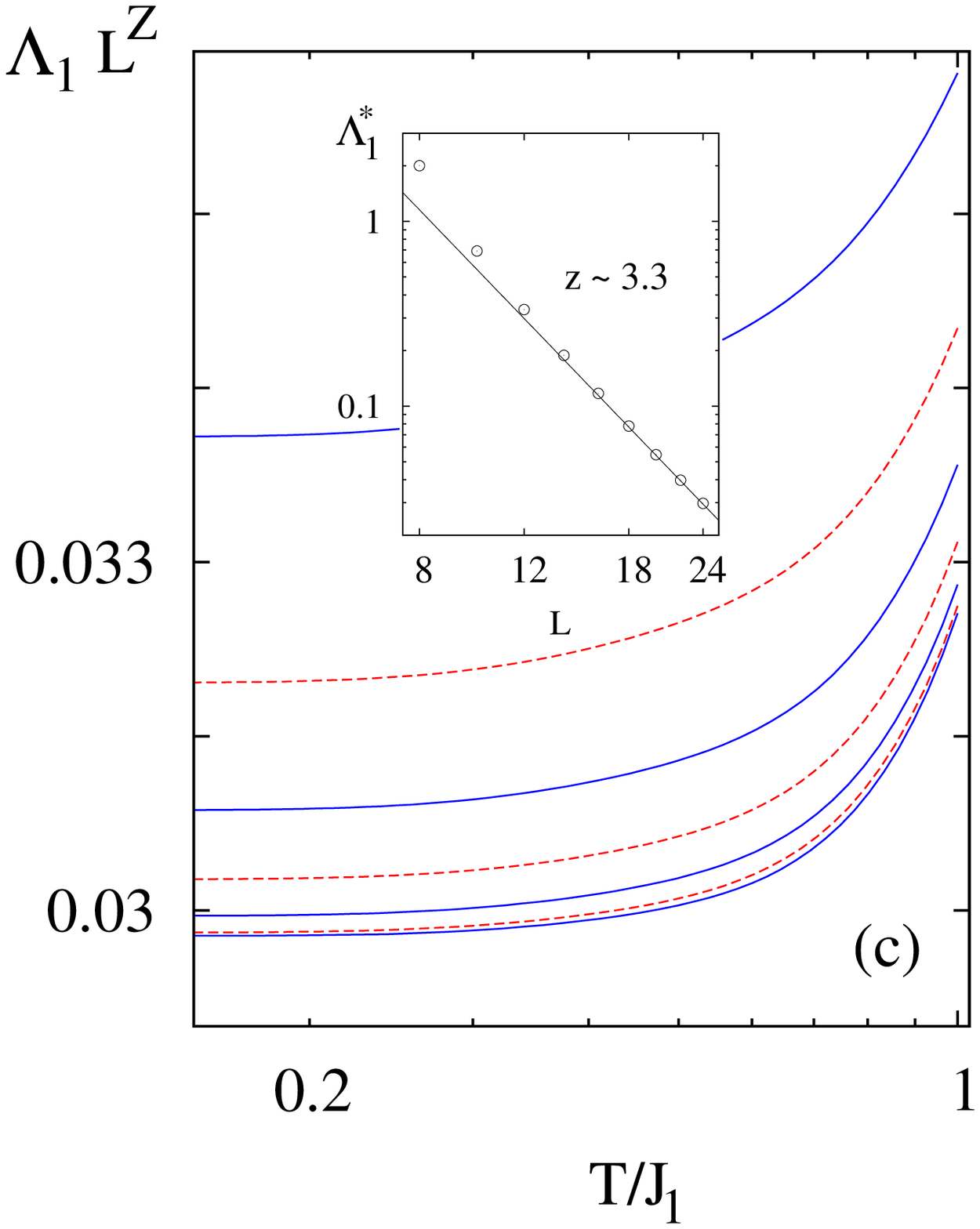}
\vskip 0.25cm
\caption{(Color online) Finite-size scaling of normalized gaps $e^{4\,(K_1+K_2)}\,
\lambda_1$ for (a) $r = 0$ (usual ferromagnetic dynamics), (b) $r = 0.3$,  and (c) 
$r = 0.7$. Solid and dashed lines show alternately the cases of $L = 12, 14, 16, 18, 20, 
22, 24$. In (a) sizes increase in upward order, while in (b) and (c) they do so from top to 
bottom. The data collapse of larger sizes was attained using dynamic exponents read off 
from the slope of the insets. These latter estimate the finite-size decay of spectral gaps 
close to $T \to 0^+$ [\,(Eq.\,(\ref{normalization})\,]. In that limit, the decay of (b) (not 
shown) is indistinguishable from that of (c) due to the common saturation trends of 
Fig.\,\ref{one}. For displaying convenience vertical scales of main panels were 
normalized by a factor $24^z$.}
\label{two}
\end{figure}

Let us start by considering the first four cases of Table~\ref{tab2}, all sharing at 
large times the decay pattern of the standard dynamics of $J_2 = 0$ \cite{details}, 
and the energy barriers $E_b =4 \,(J_1+J_2)$ alluded to in the previous section. In 
Fig.\,\ref{one} we display the above normalized gaps for several coupling ratios 
$r = J_2/J_1$ in sectors {\it (a)}, {\it (b}), and {\it (c)} in a chain of 20 sites. As 
temperature decreases, the saturated behavior of most of the $r$ values considered 
clearly signals the emergence of the expected Arrhenius regime. Note that even a 
slight deviation from the conjectured barriers would result in strong departures from 
this behavior. Also, the saturation values of $\Lambda_1$, i.e. the amplitudes involved
in Eq.\,(\ref{normalization}), come out to be $r$-\,independent so long as $r\ne 0, 1$. 
Apart from finite-size corrections (Fig.\,\ref{two}), this independence also holds for 
all other accessible lengths (a general feature applying also to other sectors of 
Table~\ref{tab2}). However in approaching $r = 0^+$ or $1^-$, where discontinuities 
already appear at the level of transition rates (see Table~\ref{tab1}), the Arrhenius 
trend is only incipient and in some limiting cases it remains beyond our reach.

When it comes to dynamic exponents ($z$), in the main panels of Fig.\,\ref{two} we 
show the finite-size behavior of these normalized gaps comparing the case of $J_2 = 0$ 
with others in sectors {\it (a)} and {\it (b)}. 
Clearly, the data collapse onto larger sizes is better for the standard dynamics, although 
in all cases the dynamic exponents producing these scaling plots are close to $z = 3$ 
(check later on the extrapolations given in Sec.\,IV\,D). In turn, their values were 
estimated from the slopes fitting the finite-size decay of $\Lambda_1$ at the saturation 
limit (insets of Fig.\,\ref{two}), this being almost identical in {\it (a)} and {\it (b)} as it 
is evidenced in Fig.\,\ref{one}.

\begin{figure}[htbp]
\vskip 0.1cm
\hskip -0.75cm
\includegraphics[width=0.56\textwidth]{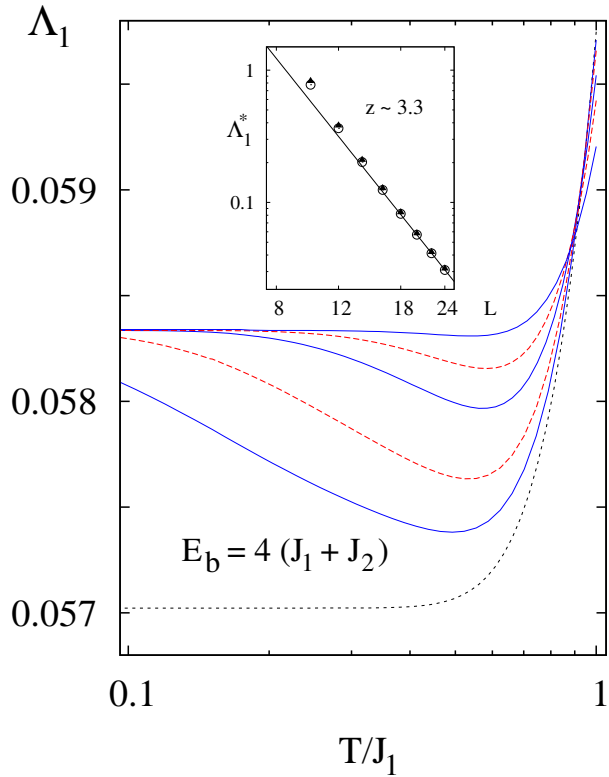}
\vskip 0.25cm
\caption{(Color online) Normalized gaps $e^{E_b/T}\, \lambda_1$ for $L = 20$ 
within regime {\it (d)} of Table~\ref{tab2}. From top to bottom solid and dashed 
lines refer in turn to coupling ratios $r = 1.5,\, 1.3, \,1.2,\, 1.1,\, 1.05$. For comparison, 
also the case  $r = 1$ [\,regime {\it (c)}\,] is shown by the dotted line. The inset 
estimates the typical finite-size decay of these gaps in the low-temperature limit 
[\,(Eq.\,(\ref{normalization})\,]. Filled triangles are representative of $r > 1$ in sector
{\it (d)}, while open circles stand for $r=1$. However, see extrapolations of Sec.\,IV\,D.}
\label{three}
\end{figure}

The above observations can also be extended to Fig.\,(\ref{three}), where the normalized 
gaps of sectors {\it (c)} and {\it (d)} are exhibited. In parallel with the difficulties 
occurring in Fig.\,(\ref{one}) as $r \to 1^-$, here these also appear in approaching the 
limit of $r= 1^+$. But otherwise, as before, the Arrhenius regime can be reached already 
within our low temperature ranges. In that respect, the inset shows that the common 
finite-size decay of $\Lambda_1^{\!^*}$ in sector {\it (d)} closely follows that of {\it 
(c)} ($r= 1$), both regimes being characterized by a slope (dynamic exponent) very 
close to that obtained for sectors {\it (a)} and {\it (b)}. Thus, an asymptotic scaling 
regime similar to the standard one of $J_2 = 0$ might be expected in these first four 
cases (despite the different proliferation rates of their corresponding M-\,states). But 
for the moment we defer that discussion to Sec.\,IV\,D. 

\subsection{$J_1 <0,\;  J_2 > 0$}

\begin{figure}[htbp]
\vskip 0.15cm
\hskip -8cm
\includegraphics[width=0.4\textwidth]{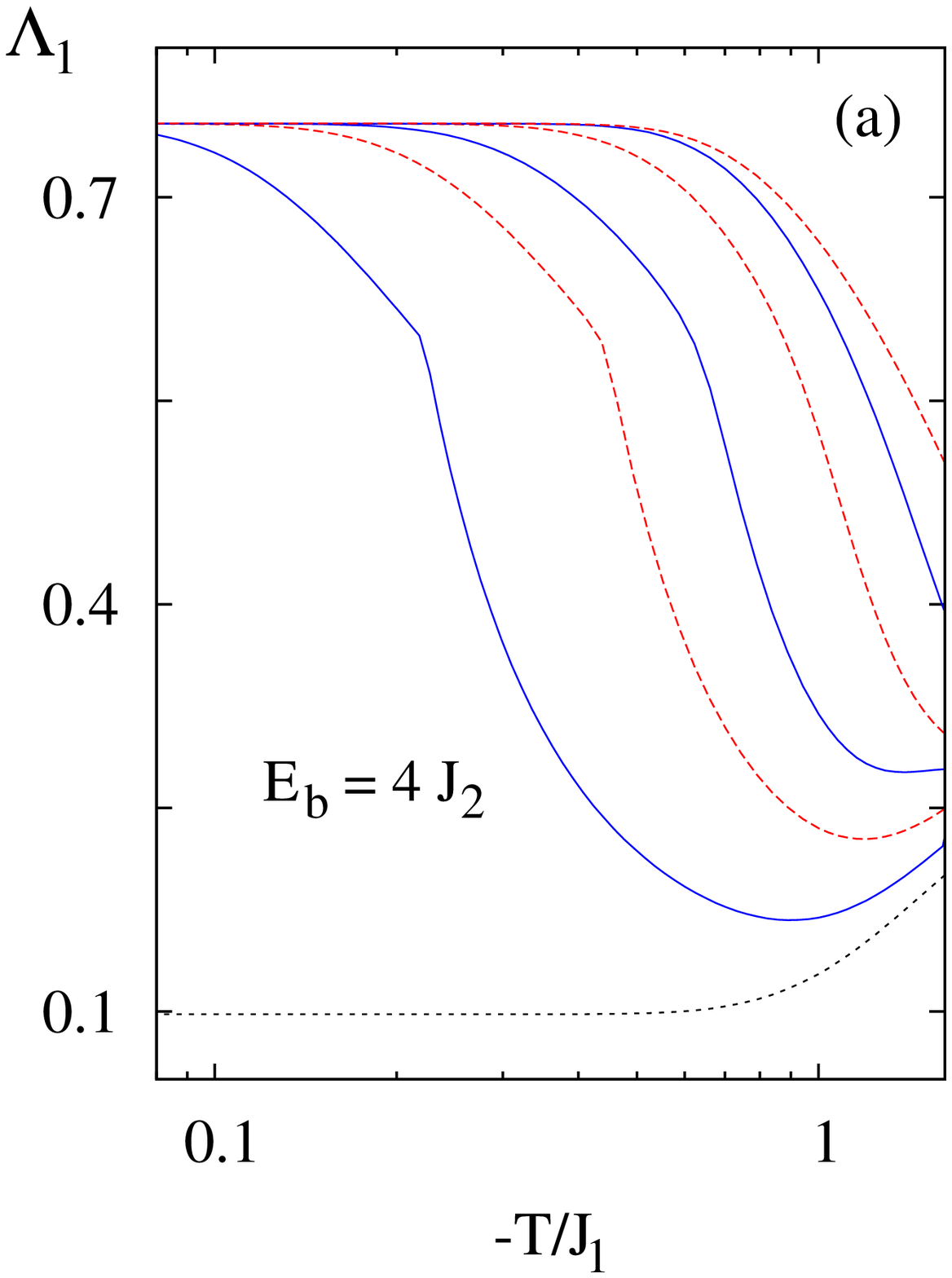}
\vskip -8.1cm
\hskip 7cm
\includegraphics[width=0.4\textwidth]{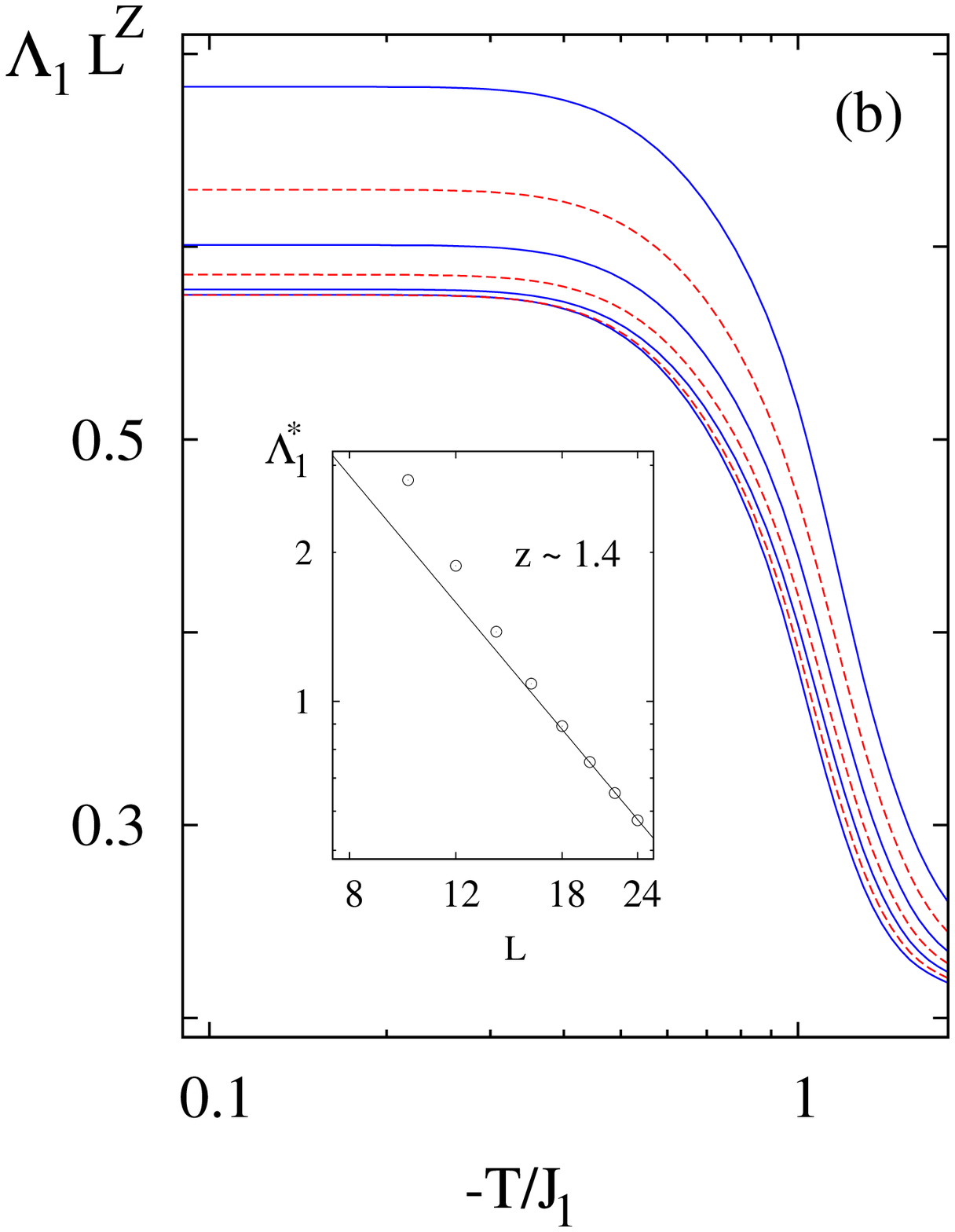}
\vskip 0.25cm
\caption{(Color online) (a) Normalized gaps $e^{E_b/T}\, \lambda_1$ in regime {\it (e)} 
of Table~\ref{tab2} using $L = 20$ . From left to right solid and dashed lines denote in
turn the cases of $-r = 0.1,\, 0.2,\, 0.3,\, 0.5,\, 0.8, 1.5$. Cusps stem from level crossings 
in the spectrum of the evolution operator. The dotted line shows for comparison the 
standard case of $r=0$ with $J_1 < 0$.\, (b) Finite-size scaling of normalized gaps for 
$r = -0.5$ upon identifying $z$ with the slope of the inset. Alternating solid and dashed 
lines in downward direction indicate sizes $L = 12, 14, 16, 18, 20, 22, 24$. For displaying 
purposes the vertical scale was normalized by a factor $24^z$. The inset estimates the 
typical finite-size decay of these gaps within sector {\it (e)} in the limit of $T \to 0^+$. 
See however extrapolations of Sec.\,IV\,D.}
\label{four}
\end{figure}

Next we turn to regime {\it (e)} where, as referred to in Sec.\,III, the dynamics follows 
a rather different decay pattern into AF states on time scales $\propto e^{4 K_2}$. As 
temperature is lowered, the saturation trends observed in Fig.\,\ref{four}(a) already 
disclose the emergence of these Arrhenius factors for several $r$-\,coupling ratios. 
As in the previous subsection, the agreement with the former is very precise given 
the persistence of the $\Lambda_1$ plateaus. Also, the amplitudes concerning 
Eq.\,(\ref{normalization}) here turn out to be $r$-\,independent although, likewise 
with what occurs in Fig.\,(\ref{one}), as $\vert r \vert$ decreases the Arrhenius 
regime barely shows up for $T/\vert  J_1\vert  \agt 0.1$. On the other hand, the trend
of decreasing minima approaching the standard non-metastable gaps of $J_2 = 0$ is  
disrupted in the low temperature limit. As discussed below, this already signals an 
abrupt crossover of scaling regimes.

Note that regardless of how small $r$ might be, the metastability of this sector does 
not disappear so long as $J_2 > 0$.  Thus, in the limit of $r =0^-$ this poses a situation 
reminiscent to that mentioned in Sec.\,I for the 1D Glauber dynamics under weak 
competing interactions. Irrespective of the weakness of the frustration, in the limit of 
$T \to 0^+$ metastability takes over and changes the dynamic exponents of that 
non-conserving dynamics from diffusive ($z = 2$) to almost ballistic ($z \sim 1$) 
\cite{Grynberg}. In that regard, here the  analogy goes deeper as a similar discontinuity 
in scaling regimes also appears in this (non-frustrated) sector. This is exemplified for $r 
= -0.5$ in the scaling plot of Fig.\,\ref{four}(b) where the data collapse towards larger 
sizes is attained on choosing a dynamic exponent $z \sim 1.4$, in turn read off from 
the slope of the inset. As in Sec.\,IV\,A, this latter depicts the finite-size behavior of 
normalized gaps within the common saturation regime of Fig.\,\ref{four}(a), thus 
showing a $\Lambda_1^{\!^*}$-\,decay which presumably is also representative of all 
$r \in$ {\it (e)}. Let us anticipate that the dynamic exponent arising from the finite-size 
extrapolations of these data (see Sec.\,IV\,D) also tends to a nearly ballistic value, far 
apart from the standard diffusive case of $ J_1<0$ ($z \sim 2$, see Fig.\,\ref{five} below) 
as well as from the subdiffusive one with $J_1 > 0$ [\,$z \sim 3.1$, Fig.\,\ref{two}(a)\,].

\vspace{-0.5cm}
\subsection{$J_1, \, J_2 < 0$}

\begin{figure}[htbp]
\hskip -8.2cm
\includegraphics[width=0.4\textwidth]{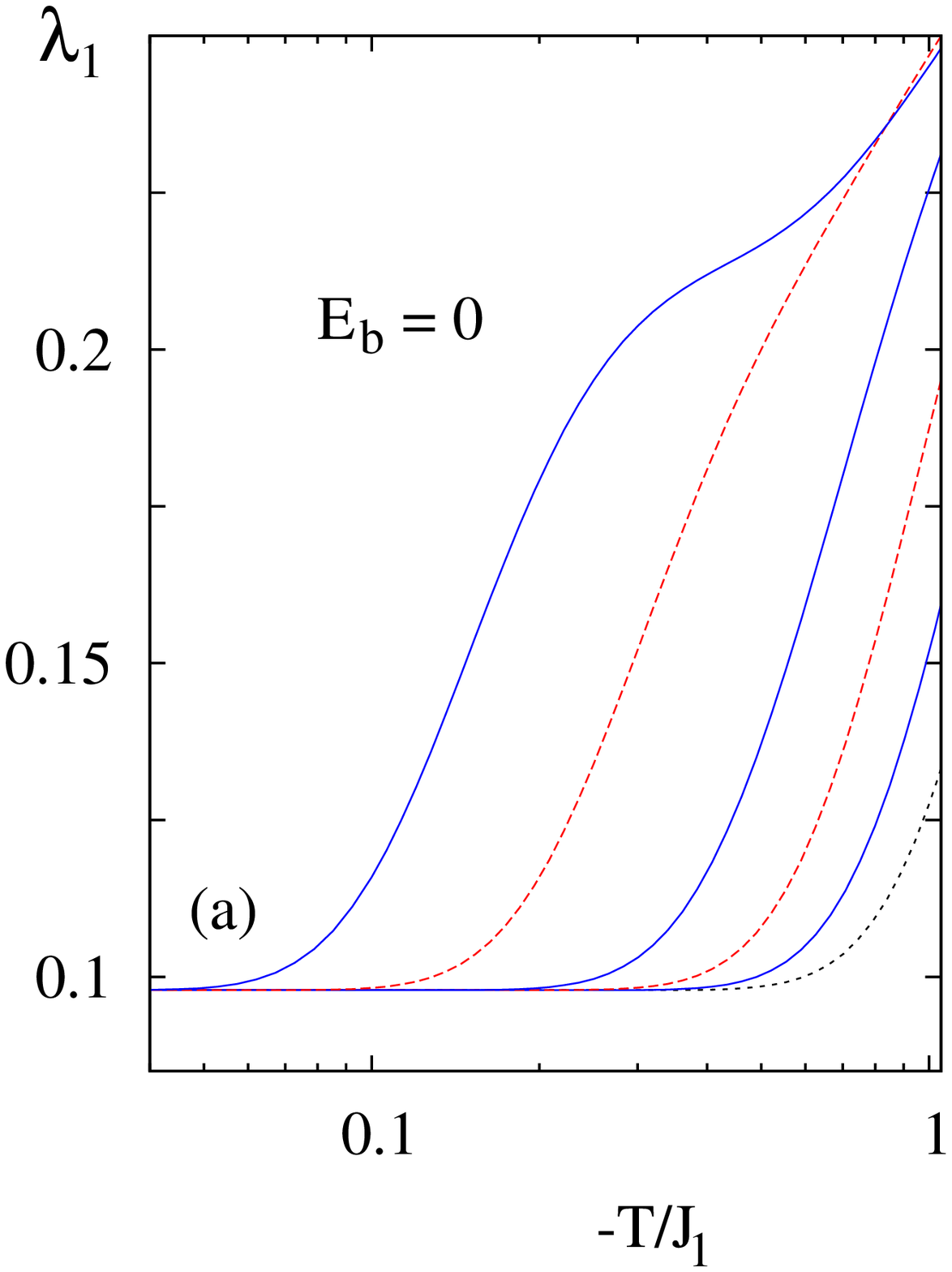}
\vskip -7.87cm
\hskip 7cm
\includegraphics[width=0.4\textwidth]{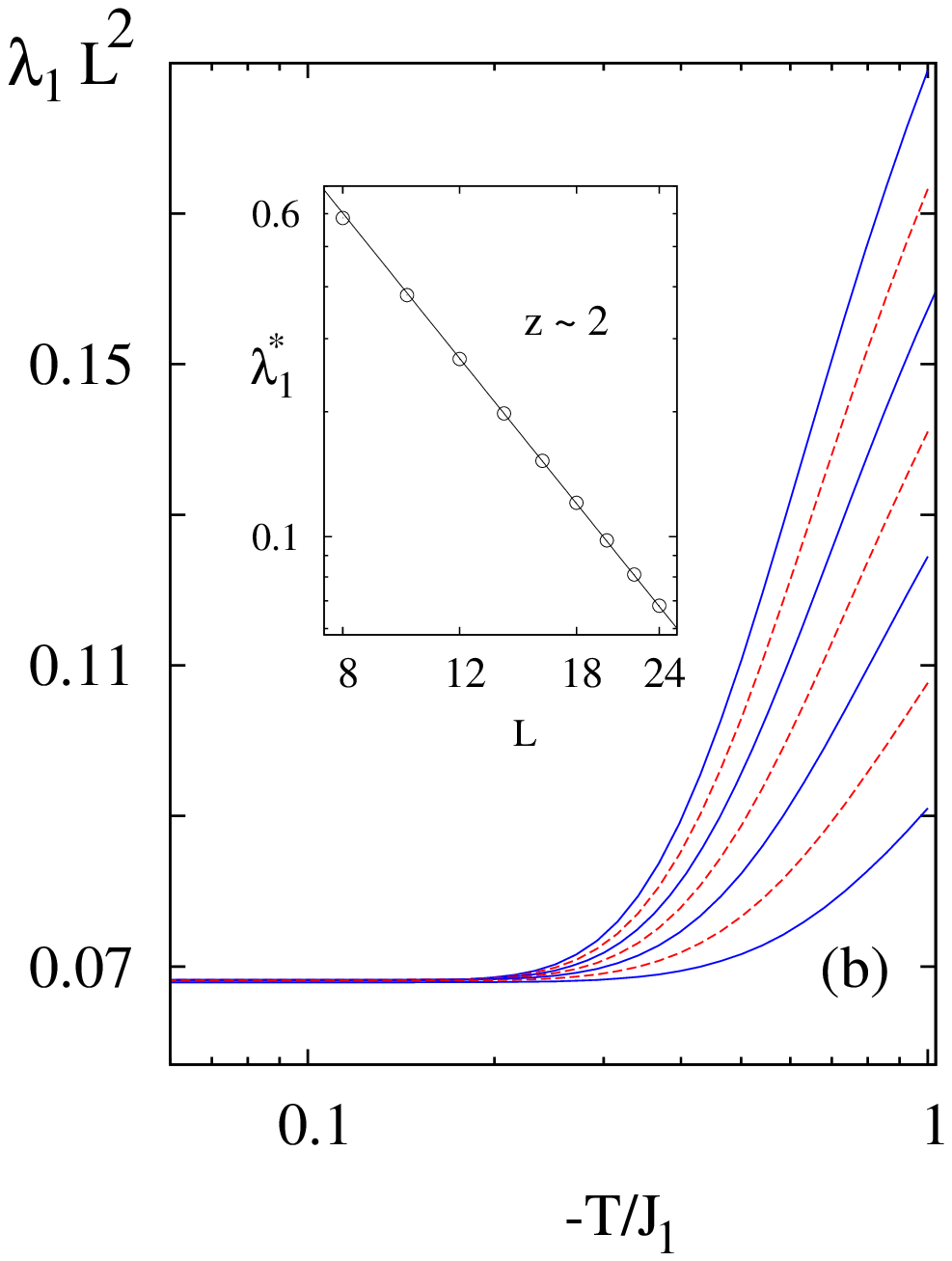}
\vskip 0.25cm
\caption{(Color online) (a) First excitation level of the evolution operator for $J_1, \, J_2 
< 0$ and $L =20$ in the non-metastable region $0 \le r  < 1/2$. From left to right solid 
and dashed lines refer alternately to coupling ratios $r =  0.45,\, 0.4,\, 0.3,\, 0.2,\, 0.1$.
The case $r = 0$ is also shown for comparison (doted line). \,(b) Scaling plot of these
levels for $r = 0.3$ and $L = 24, 22, 20, 18, 16, 14, 12$ (solid and dashed lines from top 
to bottom) using $z=2$.  For convenience the vertical scale was normalized by a factor 
$24^2$. The slope of the inset corroborates a common diffusive decay in the 
low-temperature limit of all ratios considered in panel (a).}
\label{five}
\end{figure}

Before moving on to other sectors of Table~\ref{tab2}, first we consider the 
non-metastable regime mentioned by the end of Sec.\,III, namely the situation of 
$J_1 < 0$ with $0 \le  r< 1/2$. As in sector {\it (e)}, here the phase ordering is still
AF. In Fig.\,\ref{five}(a) we display the plain spectral gaps for several coupling ratios 
in this region [\,no need of normalization as in Eq.\,(\ref{normalization})\,].
Contrariwise to all other sectors, in this case the relaxation times ($1/\lambda_1$) of 
finite chains are kept bounded in the low-temperature limit, and so the Lanczos 
convergence is now faster. Unlike the Glauber case briefly touched upon in Sec.\,IV\,B, 
here the presence of frustration does not bring about changes in scaling regimes. Taking 
for instance $r = 0.3$, this is checked in Fig.\,\ref{five}(b) where at low-temperatures 
all finite-size data can be made to collapse into a single curve by choosing the same 
diffusive exponent of the standard AF dynamics. In turn, the inset also corroborates this
by estimating the slope with which these gaps decay with the system size as $T \to 0$. 
Since in that limit $\lambda_1$ becomes $r$-\,independent (just as do the amplitudes 
accompanying the Arrhenius factors in the above subsections), clearly this scaling 
behavior persists through the entire non-metastable region.

\begin{figure}[htbp]
\hskip -7.5cm
\includegraphics[width=0.37\textwidth]{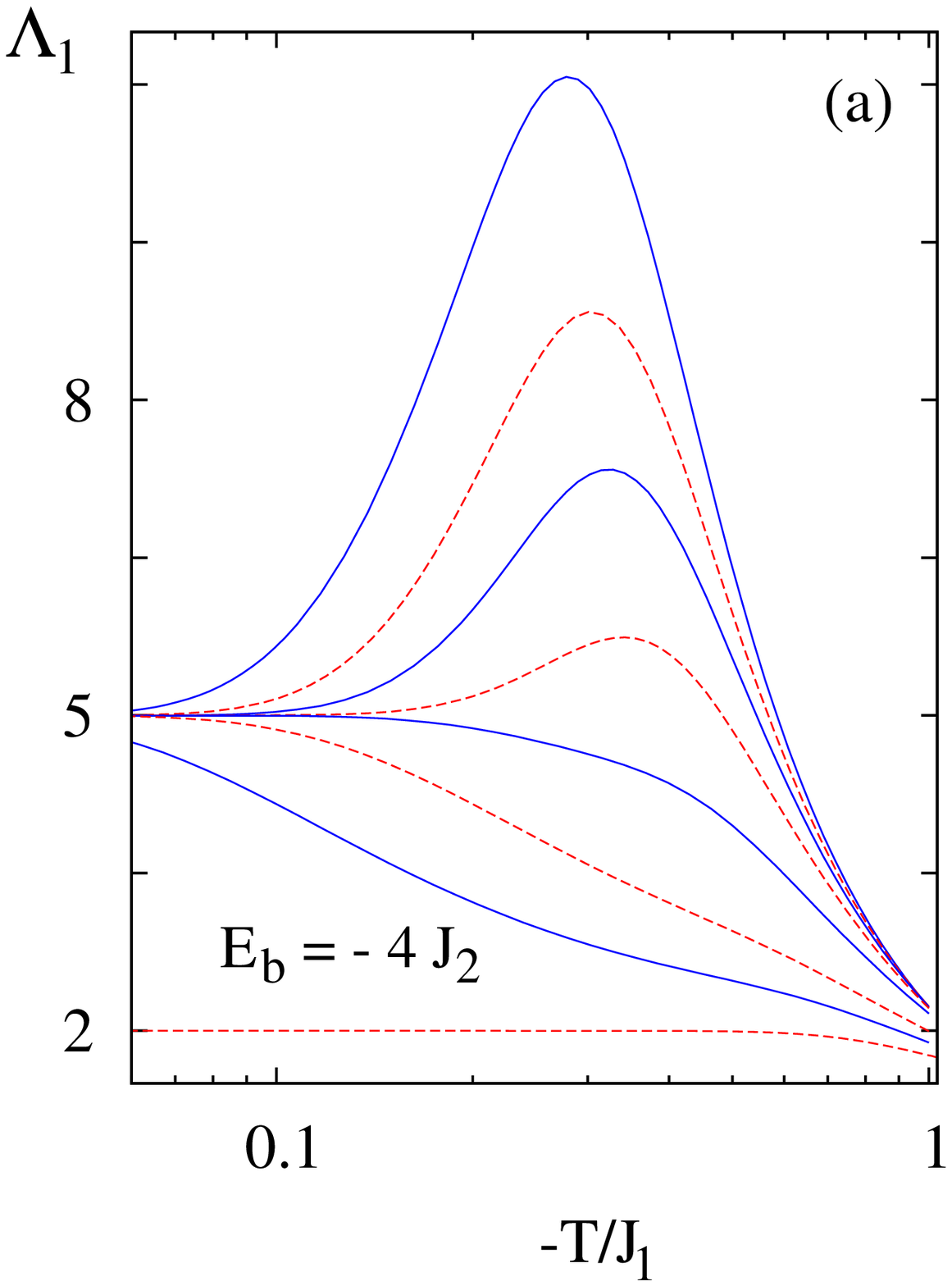}
\vskip -7.98cm
\hskip 7.2cm
\includegraphics[width=0.395\textwidth]{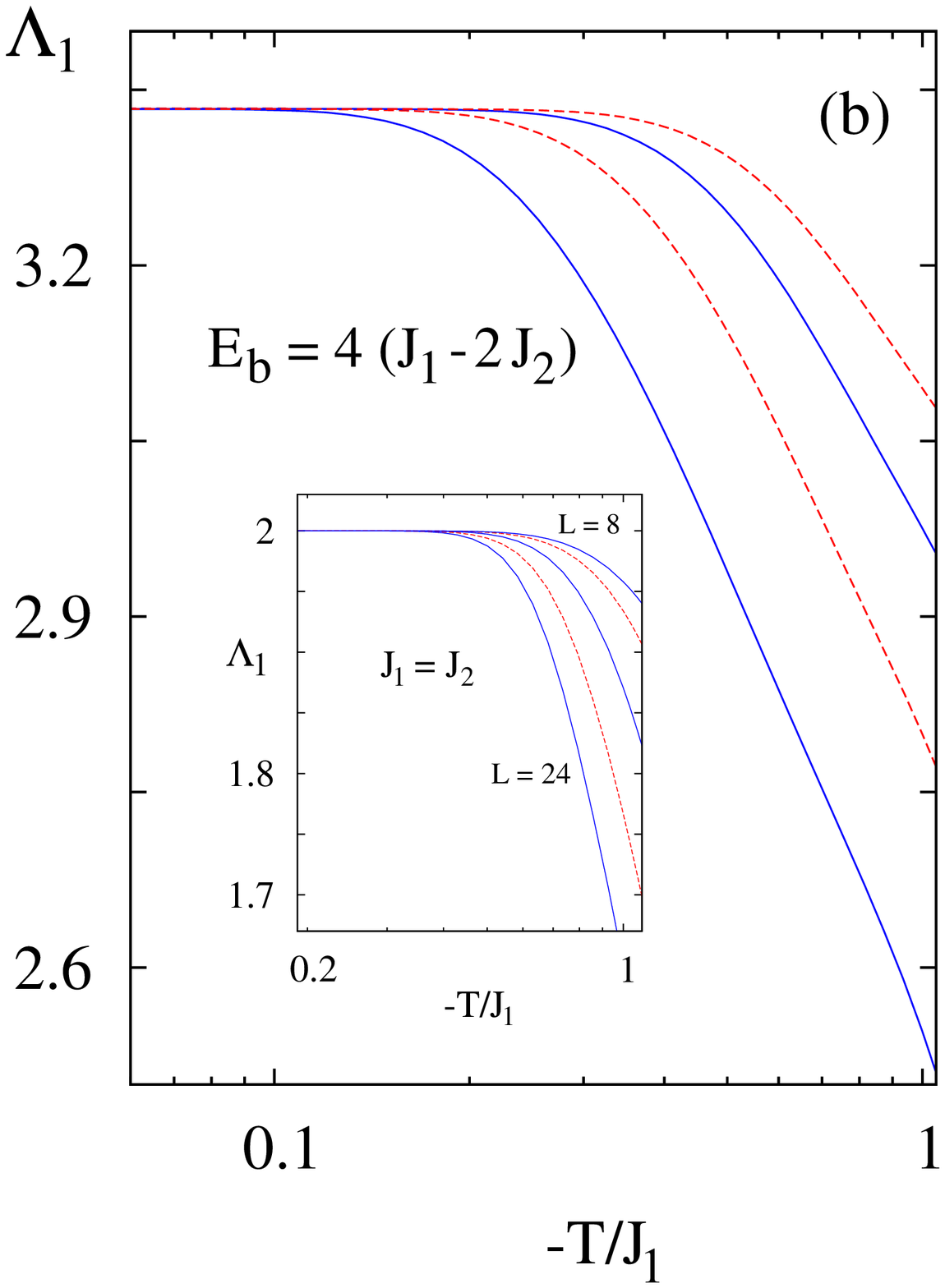}
\vskip 0.25cm
\caption{(Color online) Normalized gaps $e^{E_b/T}\,\lambda_1$ for $L = 20$ in 
sectors {\it (f)} and {\it (g)} of Table~\ref{tab2} [\,panels (a) and (b) respectively, 
exhibiting results of several coupling ratios\,]. From top to bottom solid and dashed 
lines stand in turn for (a),  $r = 0.7,\, 0.712,\,  0.725,\,  0.75,\,  0.8,\, 0.9,\, 0.95,\, 1$, 
and (b), $r = 1.5,\, 1.4, \,1.3,\,1.2$.  The inset shows the case of $r=1$ for which these 
gaps become size-independent in nearing the Arrhenius regime. Here, lengths $L = 24, 
20,16, 12, 8$ (in upward direction) match the four-fold periodicity of the ground state 
alluded to in the text.}
\label{six}
\end{figure}

Turning to sectors {\it (f)} and {\it (g)}, there are various thermal barriers affecting the 
decay of their respective M-\,structures, namely (in increasing order) $2 Q,\, - 4 K_2\,$ 
for $r \in$ {\it (f)}, and $2 P,\, - 4 K_2,\, 2 Q\,$ for $r \in$ {\it (g)}. Among these barriers, 
it turns out that actually the largest one of each sector is comprised in the normalized 
gaps exhibited in Fig.\,\ref{six}. As before, the precision of the corresponding Arrhenius 
factors is reflected in the clear saturation behavior obtained in the low-temperature 
regime. However just as in Sec.\,IV\,A, in approaching $r = 1^-, 1^+$ or $1/2^+$ 
[\,where the trend of increasing maxima continues for heights larger than the range of 
Fig.\,\ref{six}(a)\,], the discontinuities arising in transition rates carry that saturation 
limit beyond our reach. But surprisingly, as is shown by the inset of Fig.\,\ref{six}(b), 
for $r = 1$ that limit becomes {\it size}-\,independent. This suggests an exponentially 
fast relaxation even in the thermodynamic limit, though as $T \to 0^+$ the time scales 
involved get arbitrarily large, i.e. $\tau \simeq \frac{1}{2}\,e^{- 4 K_2}$.

In considering the finite-size behavior of $\Lambda_1^{\!^*} (L) $ for other coupling 
ratios in sectors {\it (f)} and {\it (g)}, note that there the four-fold periodicity of the 
ground state $\,\cdots \bullet \bullet \circ \circ \cdots$ mentioned by the beginning of 
Sec.\,II leaves us with few sizes to draw conclusions about dynamic exponents. However, 
it is worth mentioning that the rather small logarithmic slopes resulting from the gaps 
of $L = 16, 20$ and 24 (namely, $0.18$ and 0.16), are consistent with the 
size-\,independent gaps obtained for $r = 1$.

\begin{figure}[htbp]
\vskip 0.35cm
\includegraphics[width=0.41\textwidth]{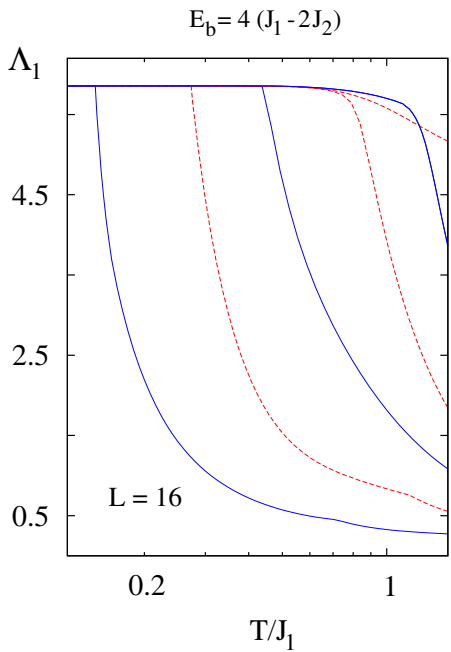}
\vskip 0.25cm
\caption{(Color online) Normalized gaps  $e^{E_b/T}\, \lambda_1$ in regime {\it (h)} of 
Table~\ref{tab2}, using $L = 16$ for $-r = 0.1,\, 0.2,\,  0.3,\, 0.4,\, 0.55,\, 1.5$ (solid and 
dashed lines from left to right).}
\label{seven}
\end{figure}

\vskip 0.25cm
{\it Sector (h)}.--\, As before, there are several thermal barriers affecting the M-\,states 
of this sector ($- 4K_2,  4K_1,  2 P,  2Q$), though now the largest one ($2 Q$) ends up 
imposing even more severe restrictions on the Lanczos procedure as temperature is 
lowered. In fact, for $L > 16$ the relenting convergence pace precluded us to obtain 
further results within the Arrhenius regime. In part, this also stems from level crossings 
in the spectrum of the evolution operator, on the other hand responsible for the pointed 
cusps observed in Fig.\,\ref{seven}. There, we just content with evidencing the presence 
of a common activation factor characterizing the decay towards either the ferromagnetic 
or four-fold ground state ($-1/2 < r < 0$, or $r < -1/2$ respectively).

\vspace{-0.3cm}

\subsection{Extrapolations}

Armed with the finite-size estimations of the normalized gaps evaluated in Secs.\,IV\,A 
and IV\,B, next we turn to the issue of going a step further than the scaling plots 
considered so far. In that respect, an improved estimation of dynamic exponents 
can be made by introducing the sequence of approximants or effective exponents 
\begin{equation}
\label{extra-Z}
Z_L = \frac{\ln[\,\Lambda_1^{\!^*} (L) / \Lambda_1^{\!^*} (L\!-\!2)\,]}
{\ln[\,(L\!-\!2)/L\,]}\,,
\end{equation}
each of which simply derives a measure of $z$ from the gaps of successive chain lengths. 
Similarly, it is worth introducing a sequence of approximants to the amplitudes involved 
in Eq.\,(\ref{normalization}), as their common saturation values strongly suggest that 
these quantities are robust within each coupling sector of Table~\ref{tab2}. Thus, 
concurrently with Eq.\,(\ref{extra-Z}) we shall also consider the accompanying set 
of effective amplitudes $A_L$ given by
\begin{equation}
\label{extra-A}
\ln A_L =  \frac{\ln \Lambda_1^{\!^*} (L)\, \ln (L\!-\!2) \, - \, \ln \Lambda_1^{\!^*} 
(L\!-\!2)\, \ln (L)}{\ln[\,(L\!-\!2)/L\,]}\,.
\end{equation}

\begin{figure}[htbp]
\vskip 0.1cm
\hskip -9cm
\includegraphics[width=0.44\textwidth]{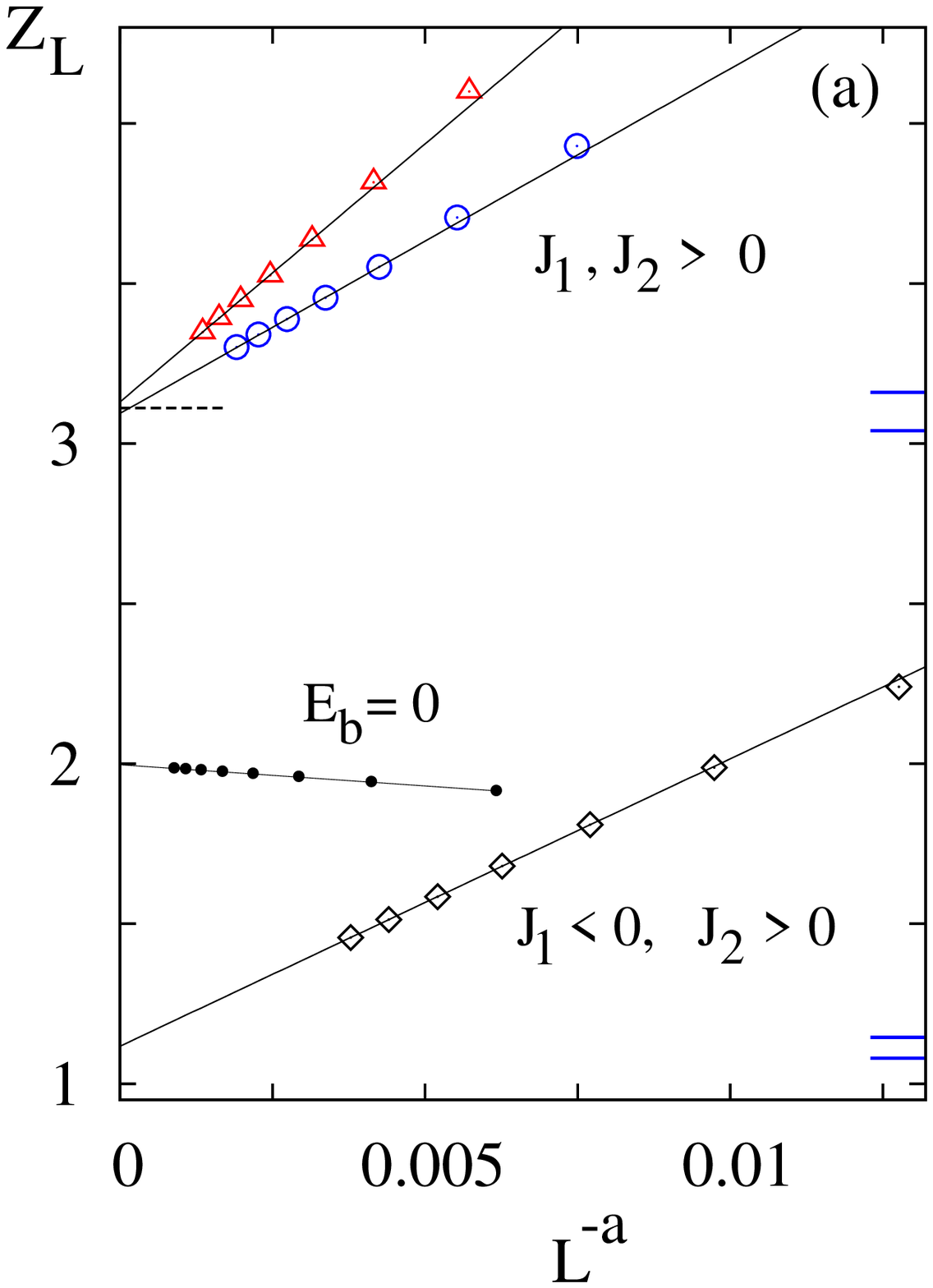}
\vskip -9.65cm
\hskip 8.5cm
\includegraphics[width=0.44\textwidth]{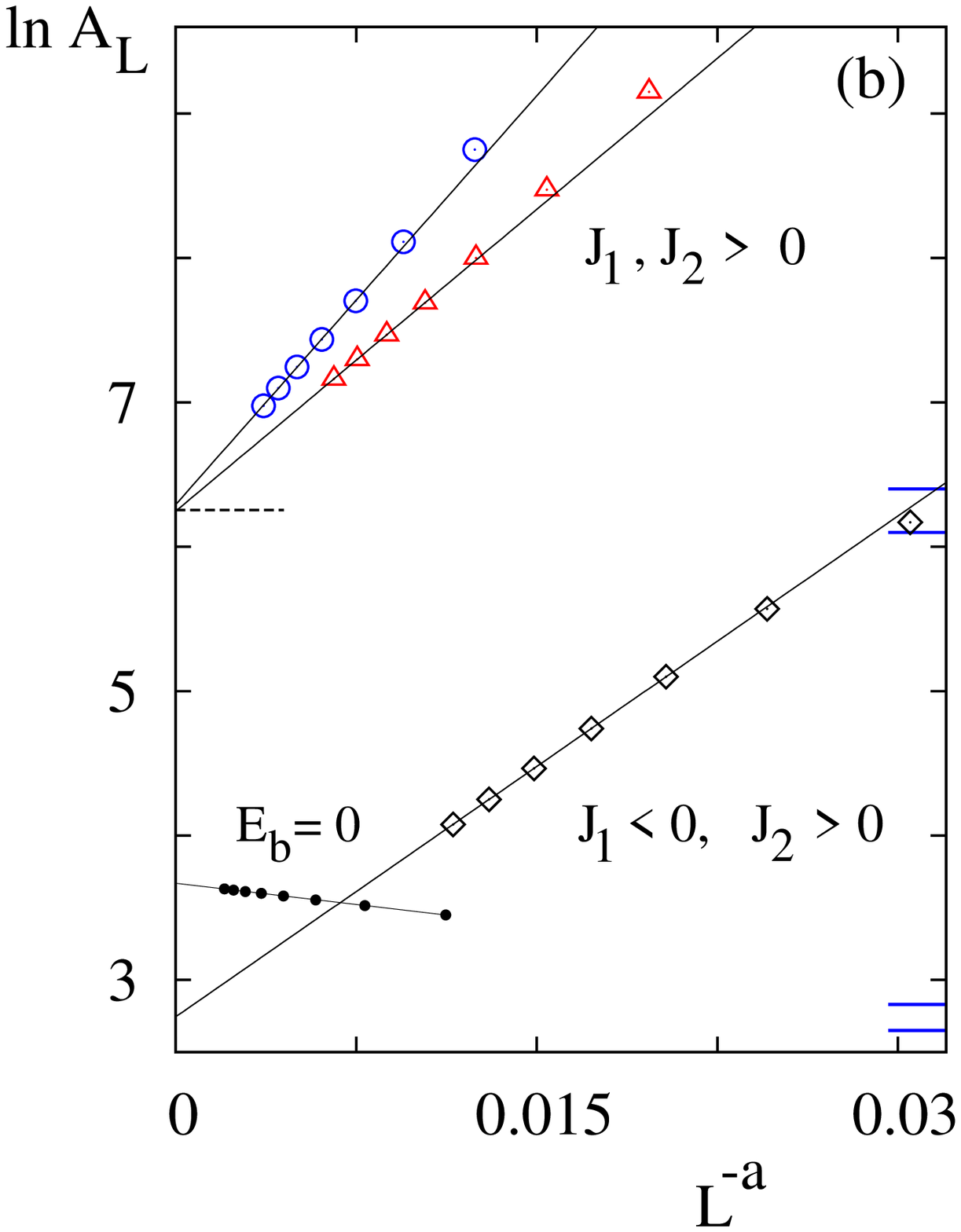}
\vskip 0.25cm
\caption{(Color online) Extrapolations of effective dynamic exponents (a), and 
amplitudes (b) defined in Eqs.\,(\ref{extra-Z}) and (\ref{extra-A}). Data of regimes 
{\it (a)} and {\it (b)} (listed in Table~\ref{tab2}), are represented by circles, regimes 
{\it (c)} and {\it (d)} by triangles, and regime {\it (e)} by rhomboids. Rightmost 
horizontal lines depict the confidence intervals arising from non-linear fittings in those 
regimes (see text for details). Leftmost dashed lines indicate the limiting values of the 
rapidly converging approximants for the usual case of $J_2 = 0,\, J_1 > 0$. Small dots 
stand for results in the non-metastable sector $0 \le r < 1/2$ with $J_1 < 0$.}
\label{eight}
\end{figure}

In general, the elements $x_L$ of a finite-size sequence obtained close to a critical point 
(here $T= 0^+$), are assumed to converge logarithmically \cite{Henkel, Guttmann} as 
$x_L = x+ \sum_j \alpha_j  L^{-a_j}$, with $\alpha$-\,constants and $a$-\,exponents 
such that $0 < a_j < a_{j+1},\;\forall\, j$. To minimize the number of fitting parameters 
here we keep only the leading-order term of that expansion which just leave us with a
nonlinear least-squares fit of three quantities. The results of those regressions are 
depicted in Fig.\,\ref{eight} which summarizes the trends of sequences (\ref{extra-Z}) 
and (\ref{extra-A}) across sectors {\it (a)} to {\it (e)}, together with those found in the 
non-metastable region. Specifically, the extrapolated exponents and amplitudes of each 
case turn out to be
\begin{equation}
\label{extrapol}
z,\; \ln A \simeq 
\begin{cases}
\,3.09(5),\;\; 6.3(1)\,,\;\;\;\, {\rm for}\,\, J_1, J_2 \in\, {\rm {\it (a)} \,\, or \,\, {\it (b)}}\,,
\vspace{0.1cm} \cr
\,3.13(3),\;\; 6.2(1)\,,\;\;\;\, {\rm for}\,\, J_1, J_2 \in\, {\rm {\it (c)} \,\, or \,\, {\it (d)}}\,,
\vspace{0.1cm} \cr
\,1.11(3),\;\; 2.74(9)\,,\;\; {\rm for}\,\, J_1, J_2 \in\, {\rm {\it (e)}}\,,
\vspace{0.1cm} \cr
\,1.996(2) ,\, 3.67(1),\,\,\; {\rm for}\,\,0 \le r < 1/2,\; J_1 < 0\,.
\end{cases}
\end{equation}
The pace of convergence of effective exponents in sectors {\it (a)} and {\it (b)} comes 
out slightly slower than that arising in {\it (c)} and {\it (d)} ($a \sim 1.97$ and 2.07 
respectively), though in the case of effective amplitudes  that pace is inverted ($a \sim 
1.76$ and 1.58). In sector {\it (e)} the convergence is still a bit slower [\,$a \sim 1.75$
in Fig.\,\ref{eight}(a) and $\sim 1.4$ in Fig.\,\ref{eight}(b)\,], but as anticipated in 
Sec.\,IV\,B the extrapolated dynamic exponent is close to that resulting from the 1D 
Glauber dynamics under weak competing interactions \cite{Grynberg}. Thus, both 
scenarios are characterized by a discontinuous crossover from a non-metastable diffusive 
regime to a metastable one with nearly ballistic exponents. In the former case ($E_b=0$) 
the convergence is somewhat faster ($a \sim 2.2$ for exponents, and $\sim 1.97$ for 
amplitudes) and the estimated errors become smaller than 0.3\%. By contrast, in sectors 
{\it (a)} to {\it (d)} the errors are such that the resulting confidence intervals superpose 
each other [\,see Eq.\,(\ref{extrapol})\,], though this is also due to the slight differences 
between the extrapolated values obtained in those sectors. Since in practice it is never 
really clear whether the assumed asymptotic behavior is sufficiently well realized by the 
data available \cite{Henkel}, those differences might well be ascribed to our finite-size 
limitations. In that sense, the merging of confidence intervals [\,rightmost center of 
panels \ref{eight}(a) and \ref{eight}(b)\,] suggests a common characterization of these 
four sectors (as anticipated by the end of Sec.\,IV\,A), within an error margin of less 
than $\sim 5 \%$. On this particular, it is also worth pointing out that for the standard 
ferromagnetic case of $J_2 = 0$ the differences between our higher approximants, 
namely $Z_{24} - Z_{22}$ and $\ln (A_{24}/ A_{22})$, are both less than 0.02\% 
[\,also see Fig.\,\ref{two}(a)\,]. The corresponding sequences approach swiftly towards 
$z \simeq\!3.11$\, (thus suggesting a slightly slower kinetics than the Lifschitz-Slyozov 
type \cite{Huse}\,), and $\ln A \simeq 6.25$, both values being consistent with 
Eq.\,(\ref{extrapol}) and included within the merged intervals of Fig.\,\ref{eight}.

Finally, we should add that the seemingly fast convergence of approximants 
(\ref{extra-Z}) and (\ref{extra-A}) in Fig.\,\ref{eight} only occurs within a small region 
of our scaled sizes ($1/L^a$,\, $a > 1$). This is due to the big $\alpha$-\,slopes stemming 
from our nonlinear least-squares fits, so that the measure of successive errors is actually 
$\alpha/L^a$. Nonetheless, the larger extrapolated errors of Eq.\,(\ref{extrapol}) resulted 
in less than 4\%. 

\section{Concluding remarks}

To summarize, we have studied a 1D Kawasaki dynamics considering up to second 
neighbor interactions thus uncovering a range of metastable situations [\,sectors 
{\it (a)} to {\it (h)} specified in Table~\ref{tab2}\,]. Following the thread of arguments 
given in the Glauber counterpart \cite{Grynberg}, we have constructed a quantum 
spin analogy whose `Hamiltonian' [\,Eqs.\,(\ref{diagonal}), (\ref{s-pairing}), and 
(\ref{s-diffusion})\,] played the role of the evolution operator of these processes in the 
kink representation (see Table~\ref{tab1}). The relaxation times of these former were 
then evaluated numerically in finite chains by analyzing the spectral gaps associated 
with those Hamiltonians using standard recursive methods \cite{Lanczos}. We focused 
attention on the low but nonzero-temperature regimes where magnetic domains tend to 
coarsen and relaxation times can grow arbitrarily large\, -even for finite chains-\, due to 
the activation barriers discussed through Secs.\,III and IV. The usual finite-size scaling 
hypothesis (\ref{scaling}) was then `normalized' as in Eq.\,(\ref{normalization}) so as 
to actually scale the spectral gaps of each sector within their corresponding Arrhenius 
regimes.

At time scales of order $e^{\,4\,(K_1+K_2)}$ the decay patterns of sectors {\it (a)} 
to {\it (d)} were argued to be those of the standard ferromagnetic case \cite{details},
although the proliferation of metastable states in sectors {\it (b)} and {\it (c)} turns 
out to be smaller. However, those differences appear to have no effect on the dynamic 
exponents, at least within the confidence intervals estimated in the extrapolations 
of Sec.\,IV\,D [\,Eq.\,(\ref{extrapol}) and Fig.\,\ref{eight}(a)\,]. By contrast, those 
extrapolations yielded nearly ballistic values for the exponents of sector {\it (e)}, on 
the other hand conjectured to decay through times $\propto e^{\,4 K_2}$ in a rather 
different form. Note here then the abrupt crossover of scaling regimes in passing from 
sector {\it (d)} to {\it (e)}. Also, in moving from this latter to the non-metastable region 
$0 \le J_2/ J_1< 1/2$ with $J_1 < 0$, another discontinuous change of dynamic exponents 
occurs. In the absence of activation barriers now these former become diffusive 
[\,Figs.\,\ref{five}(b) and \ref{eight}(a)\,] whilst the relaxation time of finite systems 
remains bounded even at $T = 0$. This situation is highly reminiscent of that of the 
Glauber dynamics studied in Ref.\,\cite{Grynberg} where the sudden emergence of 
metastable states under a small $J_2 < 0$ also changes these exponents from diffusive 
to nearly ballistic.

When it comes to sectors {\it (f)} and {\it (g)} the four-fold periodicity of the ground 
state mentioned in Sec.\,II  left us with few sizes to consider in Eq.\,(\ref{normalization}),
thus restricting our ability to extrapolate dynamic exponents. However for $J_1 = J_2 < 
0$, just where the activation barriers of these sectors coincide (see Table~\ref{tab2}), 
surprisingly the normalized gaps become independent of the system size in the Arrhenius 
regime [\,inset of Fig.\,\ref{six}(b)\,]. Clearly, this suggests an exponential relaxation 
to equilibrium through time scales $\propto e^{-4 K_2}$ that would persist up to the 
thermodynamic limit. In turn, this would be consistent with the small exponents 
preliminarily obtained for other coupling ratios in these sectors, but that is an issue 
requiring further investigation. Similarly, the study of sector {\it (h)} remains quite 
open given the convergence difficulties encountered in larger chains as temperature 
is decreased. Nonetheless, {\it all} sectors indicate that the amplitudes involved in 
Eq.\,(\ref{normalization}) possibly stand for piecewise-\,universal quantities. Except 
at $J_2/J_1 = 0, 1/2, 1$, where the original transition rates get discontinuous in the 
limit of $T \to 0^+$, this is evidenced by the common saturation values of normalized 
gaps observed throughout Figs.\,\ref{one}, \ref{three}, \ref{four}(a), \ref{five}(a), 
\ref{six}(a), \ref{six}(b), and \ref{seven}. As with dynamic exponents, those values 
were extrapolated to their thermodynamic limit in sectors {\it (a)} to {\it (e)} as well
as in the non-metastable region [\,Eq.\,(\ref{extrapol}) and Fig.\,\ref{eight}(b)\,].

In common to a variety of finite-size scaling studies (see e.g. Ref.\,\cite{Henkel} and 
references therein), ultimately small sized systems have been analyzed. Often, as is the 
case here, the dimensionality of the operators involved (transfer matrices, Liouvillians, 
quantum Hamiltonians) grows exponentially with the system size thus severely limiting 
the manageable length scales, even for optimized algorithms. In an attempt to avoid 
those limitations we also considered the scaling of relaxation times in larger chains 
via Monte Carlo simulations. However, due to the Arrhenius barriers the difficulties 
introduced by small temperatures in such simulations are by far more restrictive than 
those associated to the system size (recall that $\tau \propto e^{E_b/T} L^z$). In fact, 
starting from a disordered phase and quenching down to $T/\vert J_1\vert$ within the 
range of $0.1 -\,0.2$, it turned out that a significant fraction of the evolutions considered 
gets stuck in the typical metastable states of Table~\ref{tab2}, even at large times.

Finally, and with regard to a possible extension of this study, it would be interesting 
to derive the activation barriers $E_b$ quoted in that latter Table directly from the 
evolution operator $\cal H$ constructed in Sec.\,II\,A. However irrespective of the 
sector considered, note that for $T \to 0^+$ the leading-order of its diagonal terms  
[\,Eq.\,(\ref{diagonal})\,] is different from that of their non-diagonal counterparts 
[\,Eqs.\,(\ref{s-pairing}) and(\ref{s-diffusion})\,]. Thereby, the identification of an 
overall Arrhenius factor in the low-temperature limit of $\cal H$ is not evident in the kink 
representation. But in view of the universal amplitudes obtained above, one can further 
ask whether there might be a uniform spin rotation $R$ around a sector dependent axis 
such that $\!\!\underset{T \to 0^+} \lim R\, {\cal H}\, R^{-1} = e^{\,-\beta E_b}\;
{\cal\hat C}$, for some sector-wise but constant operator $\cal\hat C$. That would not 
only single out activation barriers but would also allow computational access to the  
strict limit of $T \to 0^+$ via the low-lying eigenvalues of $\cal\hat C$. Further work 
along that line is under consideration.

\section*{Acknowledgments}

We thank E. V. Albano, T. S. Grigera, and F. A. Schaposnik  for helpful discussions. 
Support from  CONICET and ANPCyT,  Argentina, under Grants No. PIP 2012--0747 
and No. PICT 2012--1724, is acknowledged.

\appendix
\section{Proliferation of metastable states}

As schematized in Table~\ref{tab2}, these metastable (M) structures are characterized 
by the restrictions imposed on the number of consecutive kinks ($k$) and vacancies 
($\rm v$) scattered throughout the chain. In turn, for each coupling sector these 
constraints affect the rate at which these configurations proliferate with the system
size. In order to evaluate such specific rates, in what follows we will construct a set of 
recursive relations for the number $M_L$ of those states on chains of generic length 
$L$. To ease the analysis, open boundary conditions (OBC) will be assumed throughout. 
Below, we address each case in turn.

\vskip 0.25cm
{\it (a)}.--\,  Since for this sector $k=1$ and $ {\rm v }\ge 1$,  it is helpful to consider 
the relation between the quantities $F_L (1)$ and $F_L (0)$ defined as the number of 
M-\,configurations of length $L$ having respectively 1 or 0 as kink occupations on their
first site. Clearly, under OBC these latter quantities must then be recursively related as 
\begin{equation}
F_L (1) = F_{L-1} (0)\,, \;\;F_L (0) = F_{L-1} (0) + F_{L-1} (1)\,. 
\end{equation}
Therefore, either of these quantities as well as the total number $M_L = F_L (0) + F_L 
(1)$ of M-\,states follow a Fibonacci recursion $M_{L+2} = M_{L+1} + M_L$, from which 
an exponential growth $M_L  \propto g^L$ with golden mean $g = (1 + \sqrt{5} )/2$ is 
obtained for large sizes (this growth also coincides with that of the ground state 
degeneracy at $-J_2 /\vert J_1\vert = \frac{1}{2}$ \cite{Gori}\,).

\vskip 0.25cm
{\it (b)}.--\, In addition to the kink restrictions of the previous case, here there is also a 
ban on sequence parts of the form $...\,1 0 1 0 1\,...$ as indicated in Table~\ref{tab2}.
To take into account that further constraint it is now convenient to introduce the number 
$G_L (n_1, n_2)$ of M-\,sequences of length $L$ having $n_1$ and $n_2$ as their first 
and second characters respectively ($n_1, n_2 =$ 0 or 1). Under OBC it is then a simple 
matter to check that these quantities must be related as
\begin{subequations}
\begin{eqnarray}
G_L (0,0) &=& G_{L-1} (0,0) + G_{L-2} (1,0)\,,\\
\label{b-b}
G_L (0,1) &=& G_{L-1} (1,0) -  G_{L-2} (0,1)\,,\\
G_L (1,0) &=& G_{L-1} (0,0) + G_{L-1} (0,1)\,,
\end{eqnarray}
\end{subequations}
while clearly $G_L (1,1) \equiv 0$. In Eq.\,(\ref{b-b}),\, $G_{L-2} (0,1)$ cancels out just 
all extra sequences from $G_{L-1} (0,1)$ which would not form part of $G_L (0,1)$.
Thereby, it can be readily verified that all $G$'s, along with the total number of 
M-\,states, i.e. $M_L = \sum_{n_1,n_2} G_L (n_1,n_2)$, will then follow the recurrence
\begin{equation}
M_{L+5} = M_{L+4} + M_{L+2} + M_L\,.
\end{equation}
The general solution of this latter \cite{Lando} is associated to the roots of the 
polynomial $x^5 - x^4 - x^2-1$, thus for long chains, where the largest root dominates, 
the M-\,configurations of this sector finally turn out to grow as $\sim 1.5701^L$. 

\vskip 0.25cm
{\it (c)}.--\, Further to $k = 1$, in this coupling sector every kink must appear separated 
by at least two vacancies, i.e. ${\rm v \ge 2}$, so now there are even more reductions 
in the number of M-\,states. On considering for instance the $F_L (0)$ and $F_L (1)$ 
quantities referred to in case {\it (a)}, it is clear that under OBC here these should verify
\begin{equation}
F_L (0) = F_{L-1} (0) + F_{L-2} (1)\,, \;\;\; F_L (1) = F_{L-1} (0)\,,
\end{equation}
from where the total number of M-\,configurations is obtained recursively as
\begin{equation}
M_{L+3} = M_{L+2} + M_L\,.
\end{equation}
Thus, for $L \gg 1$ the largest root of the associated polynomial $x^3 - x^2 - 1$ 
implies that $M_L \propto 1.4655^L$.

\vskip 0.25cm
{\it (d)}.--\, In this case not only $\rm v \ge 2$ and $k=1$, but also there may be 
consecutive kinks now appearing in groups of $k > 3$. To evaluate the proliferation 
of the corresponding M-\,states it is convenient to reintroduce here the $G_L (n_1, n_2)$ 
quantities referred to in case {\it (b)}. For these latter, we readily obtain the recursive 
relations
\begin{eqnarray}
\nonumber
G_L (0,0) &=& G_{L-1} (0,0) + G_{L-2} (1,0) + G_{L-2} (1,1)\,,\\
G_L (1,0) &=& G_{L-1} (0,0)\,,\\
\nonumber
G_L (1,1) &=& G_{L-1} (1,1) + G_{L-3} (1,0)\,,
\end{eqnarray}
(OBC throughout), evidently now with $G_L (0,1) \equiv 0$ as there can be no isolated 
vacancies. Thus, after a small amount of algebra it turns out that the total number of 
M-\,configurations as well as all $G's$ satisfy the recursive form
\begin{equation}
M_{L+6} = 2\,M_{L+5} - M_{L+4} + M_{L+3} - M_{L+2} + M_L\,,
\end{equation}
from where the golden mean is recovered in the largest root of the associated polynomial
$x^6 - 2 x^5 + x^4 - x^3 + x^2 - 1$. Hence, analogously to sector {\it (a)}, in the 
thermodynamic limit $M_L $ proliferates as $g^L$.

\vskip 0.25cm
{\it (e)}.--\, As it was referred to in Table~\ref{tab2} for this coupling regime $k \ge 3$ 
and $\rm v \ge 2$. Thus, resorting back to the $G_L  (n_1, n_2)$ quantities considered 
above we readily find that in this sector these must be related as
\begin{eqnarray}
G_L (0,0) &=& G_{L-1} (0,0) + G_{L-2} (1,1)\,,\\
\nonumber
G_L (1,1) &=& G_{L-1} (1,1) + G_{L-3} (0,0)\,,
\end{eqnarray}
whereas $G_L (0,1) = G_L (1,0) = 0$, as neither vacancies nor kinks may appear isolated  
in this case  (OBC assumed). Thereby, it can be checked that the total number of 
M-\,states is given recursively by
\begin{equation}
M_{L+5} = 2\,M_{L+4} - M_{L+3} + M_L\,.
\end{equation}
From the largest root of the polynomial $x^5 - 2x^4 + x^3 - 1$ linked to this recurrence, 
it then follows that for large sizes $M_L$ finally grows as $\sim 1.5289^L$.

\vskip 0.25cm
{\it (f)}.--\, In this sector kinks and vacancy constraints are respectively specified by 
$k = 1,2$, and $\rm v = 1$. Therefore, in terms of the $G$-quantities introduced above 
this means that their recursion relations should now read
\begin{eqnarray}
\nonumber
G_L (0,1) &=& G_{L-1} (1,0) + G_{L-1} (1,1)\,,\\
G_L (1,0) &=& G_{L-1} (0,1) \,,\\
\nonumber
G_L (1,1) &=& G_{L-1} (1,0)\,,
\end{eqnarray}
while clearly $G_{\!L} (0,0) \equiv 0$. Hence, after simple substitutions it is found that 
each of these $G$'s, and correspondingly the total number of M-\,states, all follow the 
recursive form
\begin{equation}
M_{L+3} = M_{L+1} + M_L \,,
\end{equation}
which for large sizes is taken over by the largest root of the polynomial $x^3 - x - 1$. 
Thereby, it turns out that for this coupling regime $M_L$ grows only as fast as $\sim 
1.3247^L$. 

\vskip 0.25cm
{\it (g)}.--\, Following Table~\ref{tab2}, in this coupling regime $k = 1,2$ (as before), 
but now $\rm v \ge 1$. In addition, there is also the constraint impeding the appearance 
of sequence parts of the form $...\,0 0 1 0 0\,...$ Hence, assuming as usual OBC, the four 
$G$-\,quantities of this case must be linked recursively as
\begin{subequations}
\label{g}
\begin{eqnarray}
\label{g-a}
G_L (0,0) &=& G_{L-1} (0,0) + G_{L-1} (0,1) - G_{L-3}(0,0)\,,\\
G_L (0,1) &=&  G_{L-1} (1,0) + G_{L-1} (1,1)  \,,\\
G_L (1,0) &=&  G_{L-1} (0,0) + G_{L-1} (0,1)\,,\\
G_L (1,1) &=&  G_{L-1} (1,0)\,.
\end{eqnarray}
\end{subequations}
Due to the above restriction, and on par with case {\it (b)}, here $G_{L-3} (0,0)$ appears 
subtracting unwanted sequences which otherwise would overestimate $G_L (0,0)$ in 
Eq.\,(\ref{g-a}). It is then a straightforward matter to verify that all of the above $G$'s 
(and therefore also $M_L$), ought to comply with the relation
\begin{equation}
M_{L +6} = M_{L+5} + M_{L+4} + M_{L+1} + M_L\,.
\end{equation}
So, the characteristic polynomial associated to this latter recurrence is $x^6 - x^5 - 
x^4 - x - 1$, from where it follows that at large sizes $M_L$ should proliferate as 
$\sim 1.7437^L$.

\vskip 0.25cm
{\it (h)}.--\, Finally, in this sector kinks and vacancy restrictions remain as in the 
previous case except that the ban on the sequences referred to above is now lifted. 
Thus, recursions (\ref{g}) still hold provided Eq.\,(\ref{g-a}) is modified as 
\begin{equation}
G_L (0,0) = G_{L-1} (0,0) + G_{L-1} (0,1)\,,
\end{equation}
i.e. the cancellation of sequences contained in $G_{L-3}(0,0)$ is no longer required here. 
After that modification it can be readily checked that the recursions arising in this 
coupling regime are all of the form
\begin{equation}
M_{L+3} = M_{L+2} + M_{L+1} + M_L\,.
\end{equation}
From the largest root of $x^3 - x^2 - x - 1$, we thus find that in the  limit of large $L$ 
here $M_L$ grows as $\sim 1.8392^L$.




\begin{thebibliography}{99}
\bibitem{comment1} For $J_2 < \vert J_1 \vert$ these times are associated with 
the smallest activation barrier of this sector. Otherwise, the lead is taken by the first 
pairing process of Table~\ref{tab1}, but its effects are inconsequential on times 
$\propto e^{4 K_2}$.
\bibitem{separations} For finite magnetizations the number of separating vacancies 
may be larger. Besides, three and four consecutive kinks may also appear, but here we 
just focus on the case of $S^z = 0$.
\end{thebibliography}
\end{document}